\documentclass{egpubl}

   \SpecialIssueSubmission    

\CGFccby

\usepackage[T1]{fontenc}
\usepackage{dfadobe}  

\usepackage{cite}  
\BibtexOrBiblatex
\electronicVersion
\PrintedOrElectronic
\ifpdf \usepackage[pdftex]{graphicx} \pdfcompresslevel=9
\else \usepackage[dvips]{graphicx} \fi

\usepackage{egweblnk}

\title[Croissant Charts]%
     {Croissant Charts: Modulating the Performance of Normal Distribution Visualizations with Affordances}

\author[Submission 1004]{Submission 1004}
\author[R. Fygenson, E. Bertini \& L. M. Padilla]
{\parbox{\textwidth}{\centering} R. Fygenson$^{1}$\orcid{0000-0002-0705-9000}, E. Bertini$^{1, 2}$\orcid{0000-0002-9932-0551}, and L. M. Padilla $^{1}$\orcid{0000-0001-9251-5279}
\\       
{\parbox{\textwidth}{\centering $^1$Khoury College of Computer Sciences, Northeastern University, Boston, MA, USA\\
$^2$College of Arts, Media and Design, Northeastern University, Boston, MA, USA}
}
}

\newcommand{\revi}[1]{\textcolor{black}{#1}}
  \newcommand{\rev}[1]{\textcolor{black}{#1}}
\newcommand{\pdf}[1]{\textcolor[HTML]{b5a50e}{\textbf{#1}}}
\newcommand{\qdp}[1]{\textcolor[HTML]{D1485B}{\textbf{#1}}}
\newcommand{\cten}[1]{\textcolor[HTML]{6B97A6}{\textbf{#1}}}
\newcommand{\ctw}[1]{\textcolor[HTML]{026494}{\textbf{#1}}}
\usepackage{float}
\usepackage{booktabs}
\usepackage{color,soul}
\usepackage[table, xcdraw]{xcolor}
\usepackage{amsmath}

\begin{document}

\maketitle
\begin{abstract}
   Affordances, originating in psychology, describe how an object’s design influences the physical and cognitive actions users may take. \rev{Past work applied affordance theory to visualization to explain how design decisions can impact the cognitive actions of visualization readers.} In this work, we demonstrate that affordances can complement effectiveness \revi{rankings} by further explaining the root causes behind visualizations’ task performance. To do so, we conduct a case study on \revi{static normal} probability density function plots, identifying their current affordances. Next, we \revi{identify} the optimal affordances for a common probability-comparison task and develop a novel affordance-driven visualization\revi{, the Croissant Chart, to support them.} We empirically validate the design's effectiveness through a preregistered study (\textit{n} = 808), demonstrating how affordances can inform predictable changes in task performance. Our findings underscore the potential for affordance-based approaches to enhance visualization effectiveness and inform future design decisions.


\end{abstract}  
\section{Introduction}
\label{sec:intro}
\textit{Affordance}, rooted in psychology, refers to the potential actions—both physical and mental—that users may take based on the design of an object \cite{hartson-1999-user-actionframework, gibson-ecological-vis-perception, Kaptelinin_aff_encyclo}. \revi{In the context of visualizations, \textit{mental actions} are the cognitive operations readers perform when interpreting a visualization, such as comparing values, grouping elements, or estimating trends. }
\textit{\rev{Cognitive} affordances}, or the relationship between the design of an object and the knowledge that is imparted upon the object’s user~\cite{Hartson-2003}, can improve understanding of \revi{visualization} design implications by complementing task precision evaluations~\cite{fygenson-cog-affordance-framework}. 
Because cognitive affordance offers an alternate lens of understanding how visualization designs impact mental actions, it can also \revi{suggest specific mental actions that} readers \revi{are likely to use}. In this paper, we demonstrate cognitive affordances\revi{' utility by using them to} connect visualization designs to readers' mental actions. \revi{This paper illustrates how} affordance-based visualization design can \revi{improve} task accuracy \revi{and} 
reveal the root cause of differences in task performance.

Although, affordances are well-studied in psychology and human\revi{-}computer interaction (HCI) (for a discussion, see~\cite{Kaptelinin_aff_encyclo}), they have only recently received attention in visualization research. 
Fygenson et al. translate affordance theory to visualization, characterizing cognitive affordances for visualization, summarizing related constructs, and proposing a framework to scaffold future research and discussion~\cite{fygenson-cog-affordance-framework}. 
We extend this work by applying cognitive affordances to visualization design and evaluation, laying the groundwork for future research.

To validate \revi{cognitive} affordances' utility, we demonstrate how they can support reasoning about likely causes of interpretation errors\revi{, such as} mathematically \textit{incorrect \revi{mental actions}}, and isolate specific elements of visualizations that are problematic. \revi{We then use} affordances to reason about design decisions that are likely to cue \textit{ideal \revi{mental actions}} for a task. 
We investigate \rev{affordances'} utility through a case study of \revi{static (i.e., non-interactive)} probability density function plots (PDFs). Prior research has identified systematic errors associated with \revi{static} PDF interpretation \cite{fygenson-padilla-pdf-scaling}, yet established methods for optimizing visualization design, such as the rankings of expressiveness and effectiveness \cite{munzner2015visualization}, have not been systematically applied to assess or improve PDFs. Moreover, established methods have not been used to assess alternative visualizations that are known to improve accuracy in probability-related tasks, such as quantile dot plots (QDPs) \cite{kay-qdps, fernandes2018-qdp-cdf}. 
QDPs’ superior performance indicates that alternative affordances may better support probabilistic reasoning than PDFs. Through an affordance-based analysis of \revi{these non-interactive} visualizations, 
we systematically examine how \rev{cognitive} affordances shape \revi{probability distribution} interpretation, offering the following key contributions:
\begin{itemize}
\item an analysis of PDF visualizations, demonstrating how an affordance\revi{s} can be used to redesign PDFs for a specific task
\item hypotheses about the underlying causes of \revi{a known} reasoning error with PDFs~\cite{fygenson-padilla-pdf-scaling} and predictions as to why these effects will not extend to QDPs
\item an affordance-based evaluation of cases where QDPs may \revi{lead to} interpretative challenges
\item a preregistered, empirical evaluation of the effectiveness of our redesigned PDFs, 
traditional PDFs, and QDPs 
\revi{that} link\revi{s} performance differences to our affordance-driven hypotheses
\end{itemize}
These contributions advance the understanding of how \rev{cognitive} affordances shape probabilistic reasoning.
\section{Background}
\label{sec:background}
\subsection{Cognitive affordances in visualization}
\label{sec:cog-aff-theory}
\textbf{Theoretical foundation}. 
Ecological psychologist J.J. Gibson first introduced affordances in the late 1970s to define the relationship between an object and its user in terms of action possibilities (i.e., ways in which an individual could use an object)~\cite{gibson-ecological-vis-perception}. Over the following decades, researchers in HCI and psychology adopted affordances to describe how the shapes of physical items imply their possible and intended uses~\cite{norman-psych-of-everyday-things, norman-1999}, and later, how digital interfaces imply possible actions (e.g., clicking a button)~\cite{Kaptelinin_aff_encyclo}. \rev{Visualization research has investigated the affordances of user interactions, studying visual cues that incite users to hover over chart elements to reveal more information~\cite{boy-interactive-vis-aff}, and examining how users interact with dynamic physical bar charts~\cite{taher2015physicalaffbarchart}. However, this work does not apply to non-interactive charts, which comprise much visualized information~\cite{boy-interactive-vis-aff}.}

 \rev{A recent subset of work investigates the non-physical affordances of visualizations, exploring how design impacts the cognitive actions of visualization readers, and thus the information that they are most likely to glean. Fygenson et al. provide an overview of this work, and name these non-physical affordances \textit{cognitive affordances}\cite{fygenson-cog-affordance-framework}. 
 They define cognitive affordances in visualization similarly to Norman’s affordance definition (i.e., all possible actions that a visualization’s design enables a reader to take)~\cite{norman-psych-of-everyday-things}, and specify that cognitive affordances are: (1) neither binary nor mutually exclusive, and (2) a factor of both a visualization's design and its reader.} \revi{Examples of cognitive actions that can be afforded include taking away messages about, and interpreting characteristics of, visualized data.}
They \revi{also} hypothesize that affordance-based evaluations can provide stronger logical reasoning for uncovering the root cause of misleading designs and suggesting successful alternatives. \revi{In this paper, we present an } 
initial validation of \revi{their} hypothesis. 

\noindent\textbf{Investigative methods.}
Within visualization, the investigation of affordances, or affordance-related concepts, has been supported via several methods (for a review, see \cite{fygenson-cog-affordance-framework}). The most popular method \rev{is a free-response question in which participants are shown a visualization and asked} open-ended questions about what they see (e.g., ``describe in a sentence
what is shown'')~\cite{quadri-doyouseewhatisee, carswell-spontaneous-interpretation, shah-graph-comprehension, zacks-tversky-bars-lines}.  This method can be modified by scoping such questions to specific tasks~\cite{xiong-afford-comparison}, 
or by restricting codes of interest when thematically coding collected responses~\cite{shah-1995-comprehending-line-graphs}. 
In our pilot studies (Sec.~\ref{sec:pilot}) and affordance evaluation (Sec.~\ref{sec:exp-design} and \ref{sec:exp-analysis}), we employ the free-response technique because of its exploratory nature and conventional coding. 

\noindent\textbf{Effectiveness rankings.}
Visualization design decisions and pedagogy are often supported by \revi{frameworks} that rank marks and channels by their ability to precisely communicate encoded data~\cite{munzner2015visualization, ware-info-vis-book, heer-bostock-repr, cleveland_mcgill_2012, cleveland-mcgills-shape-param-graphs, isenberg-sys-review-vis-techniques}. Extensions of precision-focused work investigate effectiveness from similar performance-based lenses, including accurate recall~\cite{borkin-beyond-memorability}, and minimal response time~\cite{livingston-eval-multivar-vis, isenberg-sys-review-vis-techniques}. Effectiveness \revi{rankings} are useful in evaluating common charts (e.g., bar and pie charts) and have been translated into automatic chart recommendation systems~\cite{mackinlay-auto-design, besher-feiner-autovisual}. Effectiveness \revi{rankings describe} the encodings that most precisely support a task, but offer little guidance as to the tasks that unprompted readers are most likely to complete (i.e., cognitive affordances). Past work has found that task effectiveness can correlate with visualizations' cognitive fit (i.e., how well a visualization corresponds to readers' mental representations of information)~\cite{vessey-cog-fit, vessey-cog-fit-empirical-study}, suggesting a relationship between readers' \revi{mental actions}, visualizations' designs, and effectiveness. 

\rev{The logical scaffolding of affordance-based investigations sets them apart from traditional performance evaluations by providing alternative diagnostic power. Response time and accuracy evaluations can illuminate a visualization’s poor performance and support \revi{its} redesign through frameworks such as the effectiveness and expressiveness rankings~\cite{munzner2015visualization}. Although these frameworks are useful, their guidance does not always increase visualizations' interpretability~\cite{bertini-all-chart-not-scatterplot}, or consistently change the information that readers are likely to glean. 
Cognitive affordances, on the other hand, \revi{can} provid\revi{e} logical conclusions about the underlying cause of correct or incorrect interpretations and, as we show in this paper, motivat\revi{e} redesigns that correlate with improved performance.} 

\noindent \textbf{Affording \revi{mental actions}.}
Low-level tasks, such as comparing the lengths of two shapes or reading an axis label, contribute to higher-level takeaways and decision-making~\cite{bongshin-lee-task-taxonomy, padilla2018decision}. \revi{By describing the probable} lower-level mental actions of \revi{visualization readers, cognitive affordances can shed light on the driving factors behind readers' interpretations.} For example, past research has investigated how bar chart affordances can inform the likelihood of comparative takeaways~\cite{fygenson2023affordances, xiong-afford-comparison, xiong-grouping-cues}, finding \revi{bar} placement can significantly change the comparisons and higher-level takeaways that readers report. Afforded cognitive actions can also drive readers' \revi{mental actions} when they encounter an unfamiliar graph. For example, if a reader sees a PDF for the first time, its continuous edge and aesthetic similarity to a line chart \revi{(see Fig.~\ref{fig:pdf-aff}, top)} may afford extracting height information \rev{as one would do when reading a line chart} (\revi{Fig.~\ref{fig:pdf-aff}, row a and b}). In turn, the reader might adopt a height-focused strategy and characterize the PDF by its maximum height. 

\subsection{Probability Density Visualizations}
\label{sec:prob-visualizations}
Communicating probabilities is a common requirement when disseminating experimental results~\cite{padilla2022know}, and 
probability distributions can be visualized in numerous ways
~\cite{padilla_review2022}. Box plots, and confidence, standard deviation, and histogram intervals display key statistical properties of distributions through visual representations of summary statistics. Despite their long-standing popularity in scientific communication for both experts and the general public~\cite{van2019communicating, Ross-intro-stats}, these visualization methods do not display the entire shape of a distribution, leading to a loss of statistical detail~\cite{correll2014error}. Additionally, prior research finds that interval visualizations and box plots induce errors in which readers incorrectly assume that values inside visualized boundaries are more probable than those just outside the boundaries, thereby misinterpreting encoded \revi{probability}~\cite{correll2014error, joslyn2021}. Visualizations that fully encode distributions' shape instead of only showing key statistical metrics can lead to more correct conclusions~\cite{greis-uncertainty}.

Many visualizations that display distribution shape, such as PDFs, violin, raincloud, and ridgeline plots, encode probability as the area under a curve. However, unlike other Cartesian-coordinate plots, these visualizations' y-axis values do not readily communicate useful information \cite{Ross-intro-stats}. Instead, extracting cumulative probabilities (i.e., the probability of a random variable being above or below a threshold or between two bounding thresholds) from these charts requires finding the relevant slice of the area under the curve and calculating the percentage of total area under the curve that the slice occupies. This calculation can be challenging, especially when a distribution is not uniform and the areas in question are nonpolygonal~\cite{fygenson-padilla-pdf-scaling}.

Modern visualization designs, such as quantile dot plots (QDPs)~\cite{kay-qdps} and hypothetical outcome plots (HOPs)~\cite{hullman-hops}, seek to maintain the same communication of distribution shape as area-encoded visualizations, while also increasing audience understanding through the use of visual frequency framing \rev{(i.e.,} expressing probabilities as natural frequencies). This method of communicating odds, such as ``6 out of 100 times'' rather than ``6\%'',  can improve comprehension of \revi{odds} in textual contexts\cite{cosmides-tooby-frequency-framing}. Frequency framing has inspired the discrete communication of probability as counts of marks instead of continuous areas \rev{in QDPs and HOPs}~\cite{fernandes2018-qdp-cdf, hullman-hops}. \revi{Members of the general public} are more likely to correctly interpret frequency-framed visualizations than PDFs in specific contexts \cite{kay-qdps, fernandes2018-qdp-cdf, hullman-hops, kale2020-qdp}.
Although they communicate more information than interval visualizations\cite{padilla_review2022}, these discretized visualizations still sacrifice detail in exchange for facilitating human reasoning. HOPs, for example, rely on animation, making them unsuitable for nondigital formats and requiring more time to extract precise information~\cite{kay-qdps}. QDPs trade some statistical resolution by using discrete dots to encode continuous distributions. This limitation is particularly relevant in low (20)-quantile QDPs, which prior studies suggest are more effective than higher (100)-quantile versions~\cite{kay-qdps}.

\begin{figure}[t]
\centering  
    \includegraphics[width=\linewidth, alt={The task ``Which solute, if either, has a higher probability of being present at t or less ppm in the sampled seawater?'' is presented above two normal probability density functions with different spreads and the threshold t labeled on both their x-axes. Next, we visualize strategies we observed people using to complete this task and whether these strategies lead to correct answers for equal area and equal height PDFs. The 5 strategies are 1) comparing the height of PDFs at their peaks, 2) comparing the height of PDFs at the threshold t, 3) comparing the areas under the PDFs' curves to the left of threshold t, 4) comparing the spreads of the PDFs to the left of threshold t, and 5) comparing the slopes of the PDFs at threshold t. Strategy 1 and strategy 3 render correct comparisons for only equal area PDFs. All other strategies render incorrect comparisons.}]{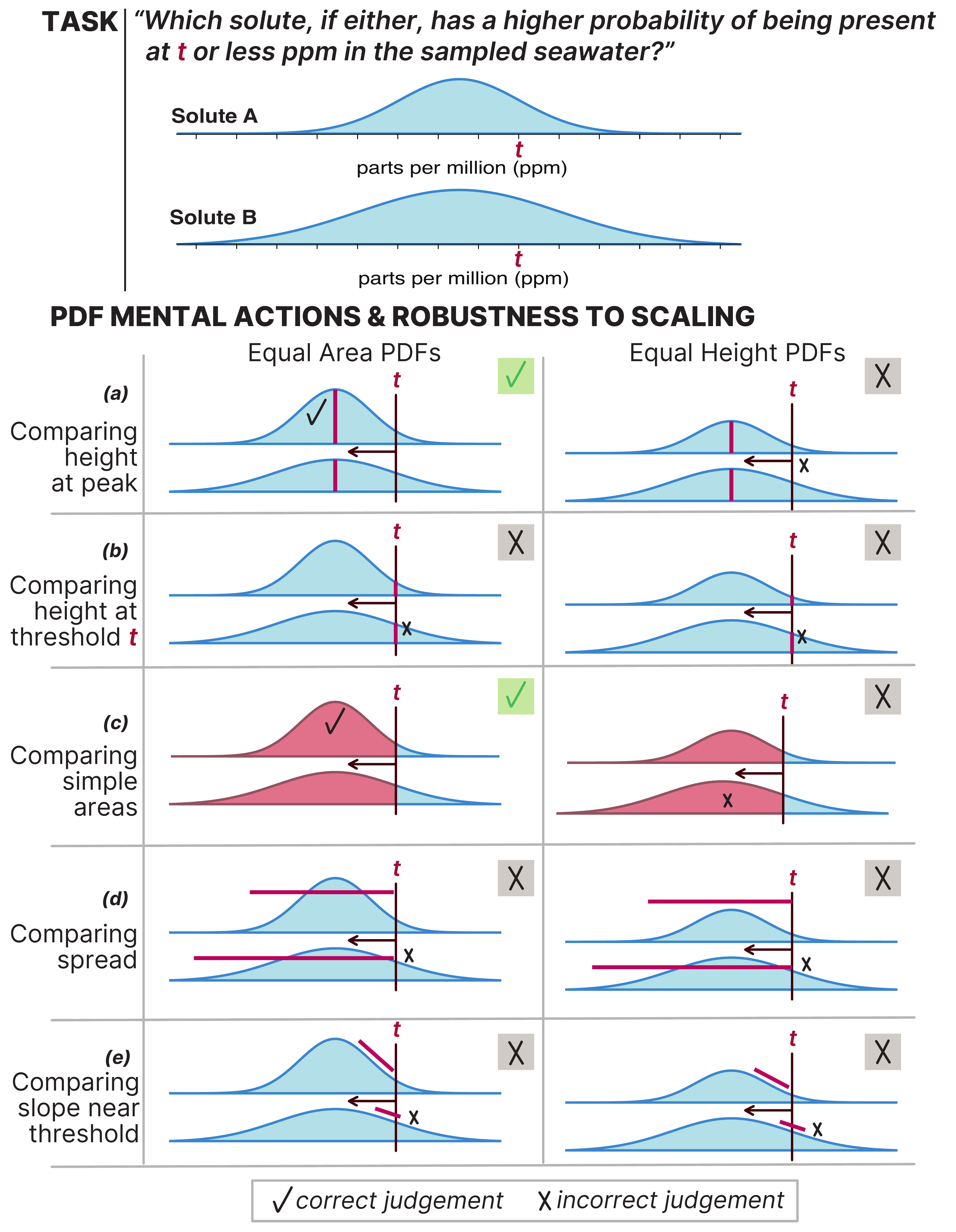}
    \vspace{-4mm}
    \caption{Top: The task we test. Bottom: PDFs' afforded \revi{mental actions} from our pilot study. Some actions yield correct results for equal-area PDFs (left), but not for equal-height PDFs (right). \revi{See Section~\ref{sec:pilot} for more detail.}}
    \label{fig:pdf-aff}
    \vspace{-8mm}
\end{figure}

\begin{figure}[t]
\centering  
    \includegraphics[width=0.9\linewidth, alt={}]{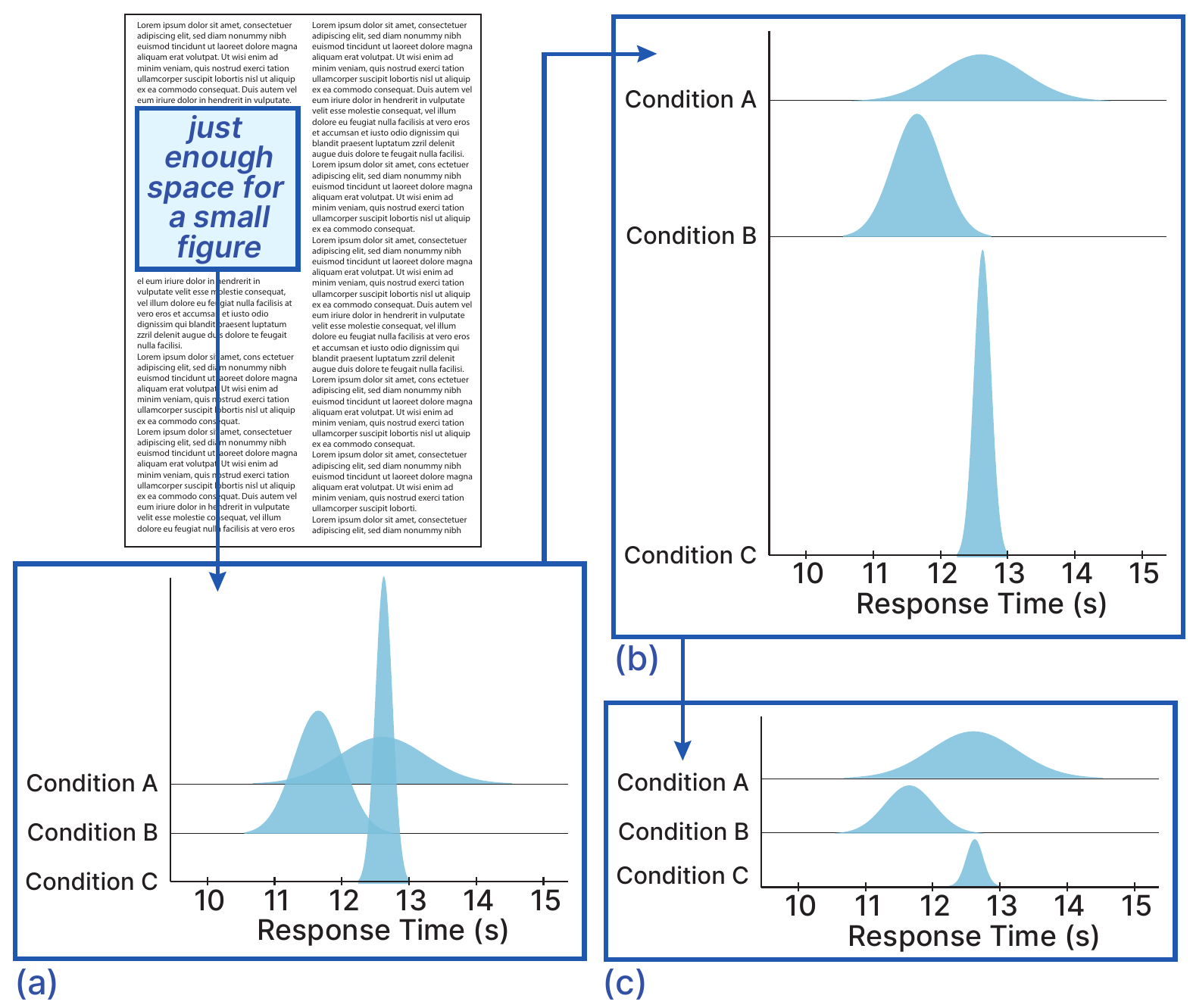}
    \vspace{-4mm}
    \caption{\revi{Why would someone scale PDFs to equal-heights? (a) different variances can cause overlap and occlusion (b) separating them requires more space (c) equal-height scaling yields a compact, statistically correct figure.}}
    \label{fig:why-squish}
    \vspace{-6mm}
\end{figure}

\noindent\textbf{Equal-height scaling vs equal-area scaling.} Statistical plotting software typically defaults to generating equal-area PDFs, such that the area under each PDF curve occupies the same number of pixels. Because each of these areas equals 1 (i.e., 100\% probability), equal-area scaling ensures all PDFs have the same probability-to-pixel ratio, allowing simple area judgments to facilitate cumulative probability comparisons (see Fig.~\ref{fig:pdf-aff}, row c). However, for area-encoded distribution visualizations (e.g., PDFs, violin plots, ridgelines), equal-area scaling can cause occlusion when some distributions are much narrower (i.e., have lower standard deviation) than others \revi{(see Fig.~\ref{fig:why-squish}, a)}. Spacing PDFs to accommodate tall distributions can use substantial vertical space \revi{(Fig.~\ref{fig:why-squish}, b) which is often restrictive to authors working within page limits, so visualization authors may choose to scale PDFs to the same height to save space (Fig.~\ref{fig:why-squish}, c).} This equal-height option is not mathematically incorrect \revi{and is popular enough to be supported by} statistical plotting software (e.g., STAN bayesplot~\cite{bayesplot-r}, ggdist~\cite{ggdist}).  Still, past research establishes that equal-height PDFs lead to more incorrect judgments than equal-area PDFs, for \revi{members of the general public} making simple cumulative probability comparisons~\cite{fygenson-padilla-pdf-scaling}. 

Notably, the impact of equal-height scaling on other probability distribution visualizations remains unexplored. However, we hypothesize that visualizations relying on frequency framing may be more robust to scaling as they encourage \revi{mental actions} independent of mark height or area. Additionally, visualizations that inherently maintain constant height, such as interval plots and box plots, fall outside this discussion, as they naturally have equal heights.

\section{Study Materials and Methods}
\label{sec:materials-methods}
Scaling pairs of PDF plots to equal heights decreases \revi{members of the general public}'s ability to compare the distributions' cumulative probability below a provided threshold (see Fig.~\ref{fig:pdf-aff})~\cite{fygenson-padilla-pdf-scaling}. Motivated by these results, we examine \revi{the utility of affordance theory}, as discussed in Section~\ref{sec:cog-aff-theory}, in 
\revi{evaluating equal-height PDFs and designing alternate visualizations}.
\revi{To do so, we proceeded in four stages. First, we analyzed pilot free-response data and relevant literature to identify possible cognitive affordances of PDFs and related distribution visualizations (Section~\ref{sec:pilot}). We assessed these cognitive affordances' alignment with the cumulative probability comparison task we investigate in our experiment. Second, we defined ideal affordances for the task and hypothesized which design features may better support them (Section~\ref{sec:redesign}). Third, we developed a novel, affordance-informed visualization to more strongly cue correct mental actions (Section~\ref{sec:redesign}). Finally, we preregistered and conducted an empirical study to test whether our predicted affordances correspond to measurable differences in task performance (Sections~\ref{sec:hyp} through~\ref{sec:exp-analysis}).}

\subsection{Affordances of probability density visualizations}
\label{sec:pilot}
\noindent\textbf{Pilot Data.} We collected pilot data to hypothesize about the affordances of different probability density visualizations by asking participants who completed simple tasks with normal probability visualizations\revi{, such as comparing the odds of two distributions,} to  ``please describe in as much detail as possible how [they] interpreted the graph and what strategies [they] used to answer the questions.'' We conducted these pilot studies via Prolific.com with participants \revi{who had Prolific approval rate over 90\%, resided in the U.S., were fluent in English, and over the age of 18 (291 female, 292 male, 14 non-binary/third gender, 4 did not report gender). These responses were between 5 and 278 words, with a median word length of 27 words.} These pilot studies allowed us to refine the measures for our main study and gain insights into afforded strategies for comparing distributions. \revi{We recruited 601 participants for the pilot study. Participants were randomly assigned to view either }normal PDFs (\textit{n} = 300) or normal QDPs (\textit{n }= 301).

We analyzed responses by thematically coding them to identify recurring concepts associated with each visualization. To do so, 
        (1) an author reviewed all responses, (2) responses were \revi{manually} grouped based on \revi{conceptual} similar \revi{in the reported} sentiments, and (3) overarching themes were identified within these groupings.    

\label{sec:pdf-aff}
\noindent\textbf{Affordances of PDFs.} To illustrate how affordance evaluations can support visualization redesigns, we selected a visualization-task use case that has been shown to lead to errors. \revi{Members of the general public} are consistently worse at comparing cumulative probabilities when using equal-height PDFs (see Fig.~\ref{fig:pdf-aff})~\cite{fygenson-padilla-pdf-scaling}. We investigated this visualization-task combination because it presents a clear judgment error, involves a sufficiently complex visualization to warrant redesign, and has alternative visualizations (e.g., QDPs) that have demonstrated high performance in other tasks. By focusing on a specific scaling issue, 
we provide a case study that exhibits the process of using affordances and their generative potential.

Our thematic coding of pilot data identified several \revi{mental actions} that likely led to incorrect calculations. \revi{Some participants} reported  comparing the heights of the normal PDFs, either at their mean (see Fig.~\ref{fig:pdf-aff} row a) or at the threshold of a cumulative probability (row b). \textit{Height-comparing} could be driven by PDFs' visual similarity to line charts \revi{(see Fig.~\ref{fig:pdf-aff} top).}
Although this \revi{action} reflects an incorrect understanding of PDFs, it is worth noting that when PDFs are scaled to equal \textit{areas}, height comparison can still result in correct comparisons (Fig.~\ref{fig:pdf-aff} row a, left). When PDFs are scaled to equal \textit{heights} (Fig.~\ref{fig:pdf-aff} row a, right), however, height comparison no longer ensures accurate cumulative probability comparisons.

Participants' also compared partial areas that correlated with the threshold
(Fig.~\ref{fig:pdf-aff} row c). \textit{Area-comparing} can be afforded by participants' interpretation of the PDF as an area chart, and although closer to a ``correct'' mathematical understanding of PDFs, also only yields correct judgments when the charts are scaled to equal areas.

Lastly, participants compared overall PDF spread (Fig.~\ref{fig:pdf-aff} row d) and slope (row e). Both \revi{mental actions} commonly led to incorrect comparisons. Participants who used PDF spread reported wider graphs having higher probabilities for our task, even though the opposite is true. Participants who used slope comparisons also fell susceptible to incorrect interpretations, such as ``... whichever curve was more steep was normally the one with less probability,'' which is the opposite of the case. 
These \textit{shape-comparing} \revi{actions} are robust to equal scaling; y-axis scaling does not change spread or significantly alter comparisons of PDF slope. Still, they led to incorrect conclusions. In summary, the first phase of our affordance evaluation identified three primary \revi{mental actions} that contribute to reasoning errors with equal-height PDFs: height-comparing, area-comparing, and shape-comparing.
\begin{figure}[h!]
\centering  
    \includegraphics[width=0.68\linewidth, alt={}]{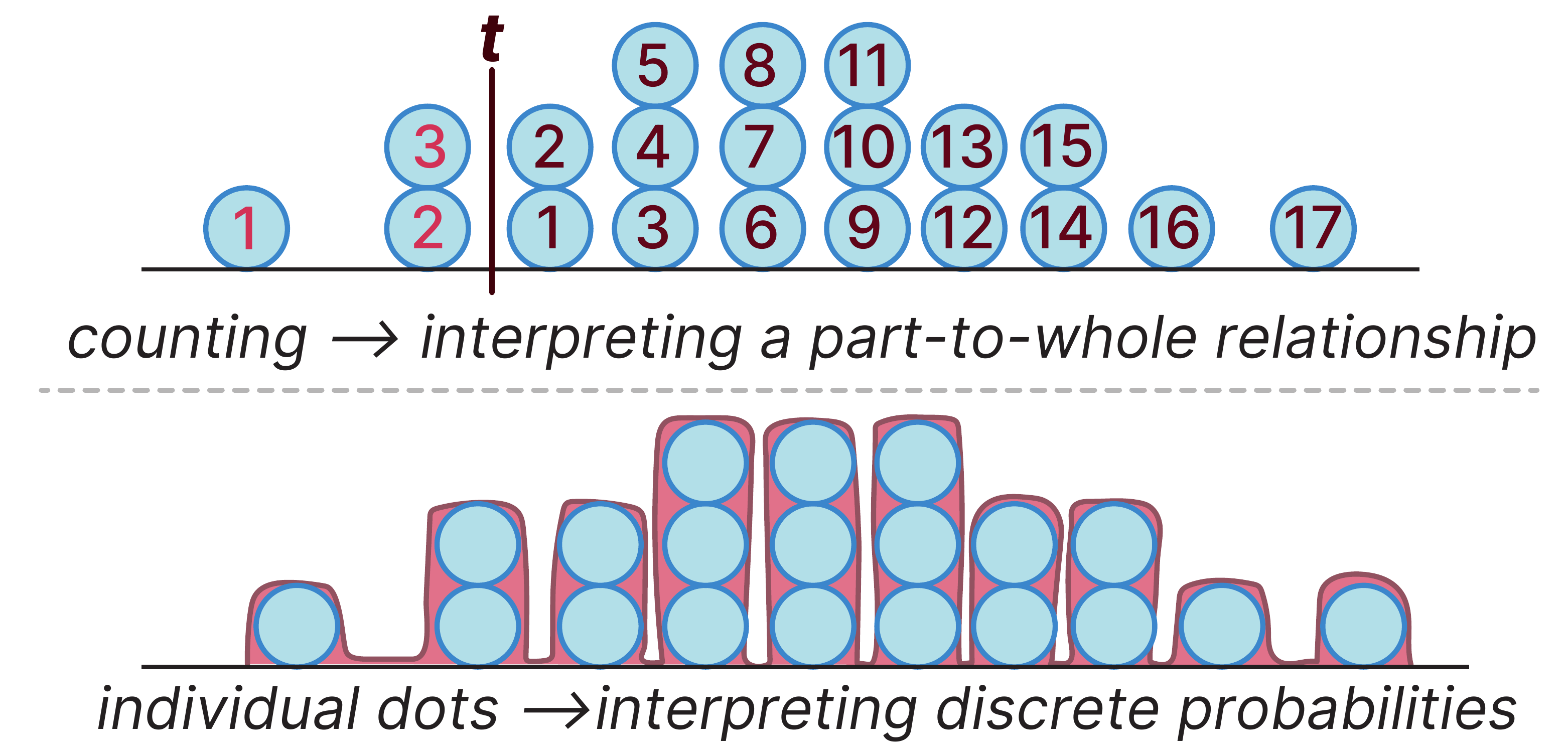}
    \vspace{-1mm}
    \caption{\revi{QDPs' afforded mental actions from our pilot study.}}
    \label{fig:qdp-aff}
    \vspace{-5mm}
\end{figure}

\noindent\textbf{Affordances of QDPs.}
A review of QDP pilot data revealed one predominant afforded strategy. Participants often reported \textit{counting} and then comparing the number of dots on the right of the cumulative threshold for each QDP \revi{(Fig.~\ref{fig:qdp-aff}, top)}.
This strategy is robust to equal-height scaling because height manipulations do not affect dot counts (see Fig. ~\ref{fig:stimuli} \revi{for equal-height QDPs}). QDPs' strong counting affordance is indicative of their frequency-framed motivation, as participants interpreted each dot as part of a greater whole (e.g., 5 dots out of 10 total).

Additionally, some participants mentioned confusion when quantile dots did not align with an x-axis value of interest, indicating that they perceived each dot as a single probability, as opposed to an approximated continuous distribution \revi{(Fig.~\ref{fig:qdp-aff})}. This discrete interpretation can lead to erroneous conclusions, especially when distributions have wide spreads that lead to large spaces between dots. This affordance aligns with past research on the nature of individual areas (e.g., bar charts) affording \revi{readers to interpret data as discrete} and connected dots (e.g., line charts) affording \revi{readers to interpret data as continuous} \cite{zacks-tversky-bars-lines, shah-graph-comprehension}. 

\noindent\textbf{Ideal affordances for equal-height PDFs.} After reviewing our pilot analysis, we identified ideal and undesired affordances from both PDFs and QDPs. Most significantly, QDPs afford counting, which makes them robust to equal-height scaling. PDFs lack this affordance, which we hypothesize contributes to incorrect judgments when scaled to equal heights (see Fig.~\ref{fig:pdf-aff}). However, QDPs also afford distributional discreteness much more than PDFs, which can lead to incorrect comparisons when two distributions share similar standard deviations (see Fig.~\ref{fig:cut-off-dist}). Thus, we identified that, for our cumulative probability task, an ideal visualization would afford both counting marks as parts of a whole and continuity. 

Next, we \revi{we constrained our design exploration to established static visualization techniques that have been empirically shown to support part-to-whole reasoning among members of the general public. Contemporary literature on part-to-whole and probability communication has strong empirical support for two approaches: icon arrays and treemaps. Icon arrays (i.e., frequency-framed displays) have been widely shown to improve probability comprehension by visually representing both the reference class and affected subset, thereby supporting accurate part–to-whole reasoning and reducing denominator neglect, especially among less numerate individuals (for reviews, see~\cite{ancker2006design, garcia2017designing}). Additionally, empirical work on rectangular treemaps suggests that spatial subdivision techniques can support accurate part–to-whole judgments. Although treemaps encode values using area, controlled experiments demonstrate judgment accuracy comparable to hierarchical bar charts at certain data densities and, in some cases, faster comparisons~\cite{kong2010perceptual}.} 
\revi{Based on these prior works,} we identified two design choices that we hypothesized afforded counting parts of a whole: 
\begin{enumerate}
    \item an ``icon array'' technique in which colored glyphs are used to signal some items out of a larger set, as is done with QDPs, and
    \item a ``spatial subdivision'' technique in which a shape is divided by thin lines into smaller shapes, as in tree maps. 
\end{enumerate}

Of these two techniques, only the latter has the potential to afford continuity. Currently, this spatial subdivision technique communicates probability distributions through standard deviation or histogram intervals by encoding quantiles or standard deviations as abutting, equal-height bars~\cite{yang2023swaying, fernandes2018-qdp-cdf}. Although these charts are robust to vertical scaling and communicate a continuous distribution across a range of values, interval visualizations are susceptible to reasoning errors (for reviews see \cite{joslyn2021, padilla_review2022}) and are less ``expressive'' than alternative visualizations, which more readily communicate details of outliers and distribution shape~\cite{padilla_review2022}. Thus, we concluded that spatial subdivision is currently employed in charts that do not show the distributional information we desire.

\subsection{An affordance-motivated redesign}
\label{sec:redesign}
\begin{figure}[h!]
\centering  
\includegraphics[width=0.75\linewidth, alt={A croissant chart with annotations to describe how each design decisions impact the chart's affordances. All annotations are described in the caption.}]{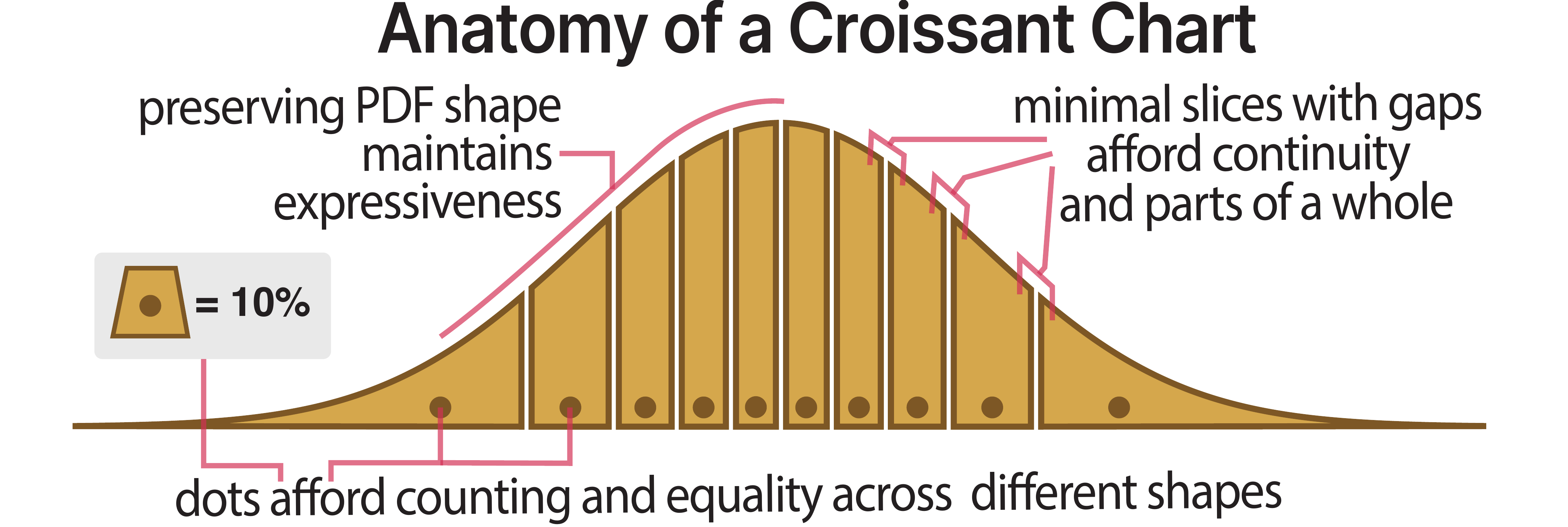}
    \caption{Croissant charts' intended affordances. PDF shape for expressiveness. Minimal slices with gaps afford continuity and parts of a whole. Dots afford counting and equality across slices.}
    \label{fig:anatomy}
    \vspace{-1mm}
\end{figure}

Above, we established our goal of affording 1) counting dots as parts of a whole, 2) continuity, and 3) a general distributional shape to maintain expressiveness. 
Below, we present an affordance-informed redesign of PDFs: \textit{Croissant charts} (see Fig.~\ref{fig:anatomy}).

\noindent\textbf{Affording counting parts of a whole.} To afford counting strategies, we used two approaches: spatial subdivision and marks that encourage counting. 
Subdivision cuts single shapes into multiple parts to communicate that the parts originate from one entity. Inspired by QDPs, we slice the PDF into quantiles, such that each slice represents an equal probability~\cite{kay-qdps}. However, we only display vertical slices to prevent the misinterpretation that horizontal slices of area under a PDF's curve lead to logical conclusions.

Additionally, we place a single dot in each quantile slice, in case QDPs' dots afford counting due to their shape. We also hypothesize that dots help mitigate the strong area, height, or width strategies we see in our pilot data on PDFs. We hypothesize that these dots correctly afford equal probabilities across each slice \rev{as a visual extrapolation of the equiprobability bias (i.e., people assume equal probabilities across items)
~\cite{gauvrit-eqprob}. 
We also use the dots to indicate equal probability in the charts' legend for added clarity. \revi{We darken the border of each section and each dot to increase their saliency at smaller scales.}}

\noindent\textbf{Affording a continuous distribution.}  To maintain perceptual continuity, we minimize area removal when subdividing the PDF. This decision draws on Gestalt principles of closure~\cite{wagemans-gestalt-perception}, and preserves as much of the PDF shape as possible. However, we include slight padding around each slice to ensure visual separation. 

\noindent\textbf{Affording general distributional shape.} To maintain the expressiveness with which PDFs display distributional shapes, we retain their density function curve rather than making each quantile the same height as in interval plots, or adopting a discretized distribution shape as in QDPs. We observed in pilot data that dots placed in QDPs' tails can skew readers' perception of a distribution's continuity and sometimes afford multimodality. Additionally, past research shows that bin width and dot size can influence the shape of QDPs~\cite{correll-prob-density2019}. By preserving the continuous shape of PDFs, we aimed for greater resilience against these distortions. 

\subsection{Investigative Questions and Hypotheses}
\label{sec:hyp}
We preregistered three main hypotheses about differences in reader performance when using PDFs, QDPs, and croissant charts on OSF. This preregistration \rev{and a visualization of our hypotheses for easier reference are available in Supplemental Materials\footnote{https://osf.io/txwj5/}.} The first hypothesis is that \textbf{(H1) equal-height PDFs will result in a lower likelihood of correct cumulative comparisons than equal-area PDFs}. This hypothesis is based on results from previous work \cite{fygenson-padilla-pdf-scaling}, and its confirmation would replicate prior findings while checking for methodological validity.

Our second \revi{set of} preregistered hypotheses examines whether visualizations that afford counting can mitigate the performance decline observed when PDFs are scaled to equal heights.
\revi{We preregistered four hypotheses concerning counting-affording designs. \textbf{Performance hypotheses:} We hypothesized that
\textbf{(H2a)} equal-height Croissant-10s,
\textbf{(H2b)} equal-height Croissant-20s, and
\textbf{(H2c)} equal-height QDPs (20 quantiles)
would each result in a higher probability of correct cumulative comparisons than equal-height PDFs.} \revi{\textbf{Robustness-to-scaling hypothesis:} We further hypothesized that
\textbf{(H2d)} QDPs, Croissant-10s, and Croissant-20s would exhibit a smaller decrease in correctness when moving from equal-area to equal-height scaling than PDFs.}



Our last experimental hypothesis is motivated by our theorizing  that QDPs only weakly afford a continuous distribution. We preregistered that \textbf{(H3) QDPs' relationship between standard deviation (SD) pair comparisons and correctness will differ from other visualizations' relationship between SD and correctness, specifically in the 4.5 vs. 5 SD condition}. In other words, QDPs' performance would drastically change in the 4.5 v 5 SD condition in a way that other visualizations' performance would not. We rationalized that when two distributions' SDs are similar, their QDPs' dots will be stacked similarly and moved only slightly along the x-axis in ways that may be too subtle for readers to recognize.

\subsection{Stimuli}
\label{sec:stimuli}
When comparing of cumulative probabilities of equal-area PDFs, readers can rely on heuristics, such as PDF height, to make accurate judgments. For equal-height PDFs, however, height comparison can lead to incorrect conclusions; other strategies, such as counting, offer a more robust approach. Following our affordance analysis of PDFs, we tested three types of visualizations: PDFs, which our pilot data suggest strongly afford continuity but only weakly afford counting, QDPs; which our pilot data suggest strongly afford counting but weakly afford distribution continuity; and croissant charts, which we designed to afford both continuity and counting.

We test QDPs with 20 quantiles because they have been shown to result in better performance than their 100-quantile counterparts \cite{kay-qdps}. We test croissant charts with 20 quantiles (Croissant-20s), as mathematical equivalents to our QDP stimuli, and croissant charts with 10 quantiles (Croissant-10s). Our pilot data showed that Croissant-10s outperform croissants with five quantiles, so we wanted to study how Croissant-10s compare to Croissant-20s.   

\begin{figure}[h]
\centering   
\includegraphics[width=\linewidth, alt={Stimuli variations. All variations and labels are described in the caption.}]{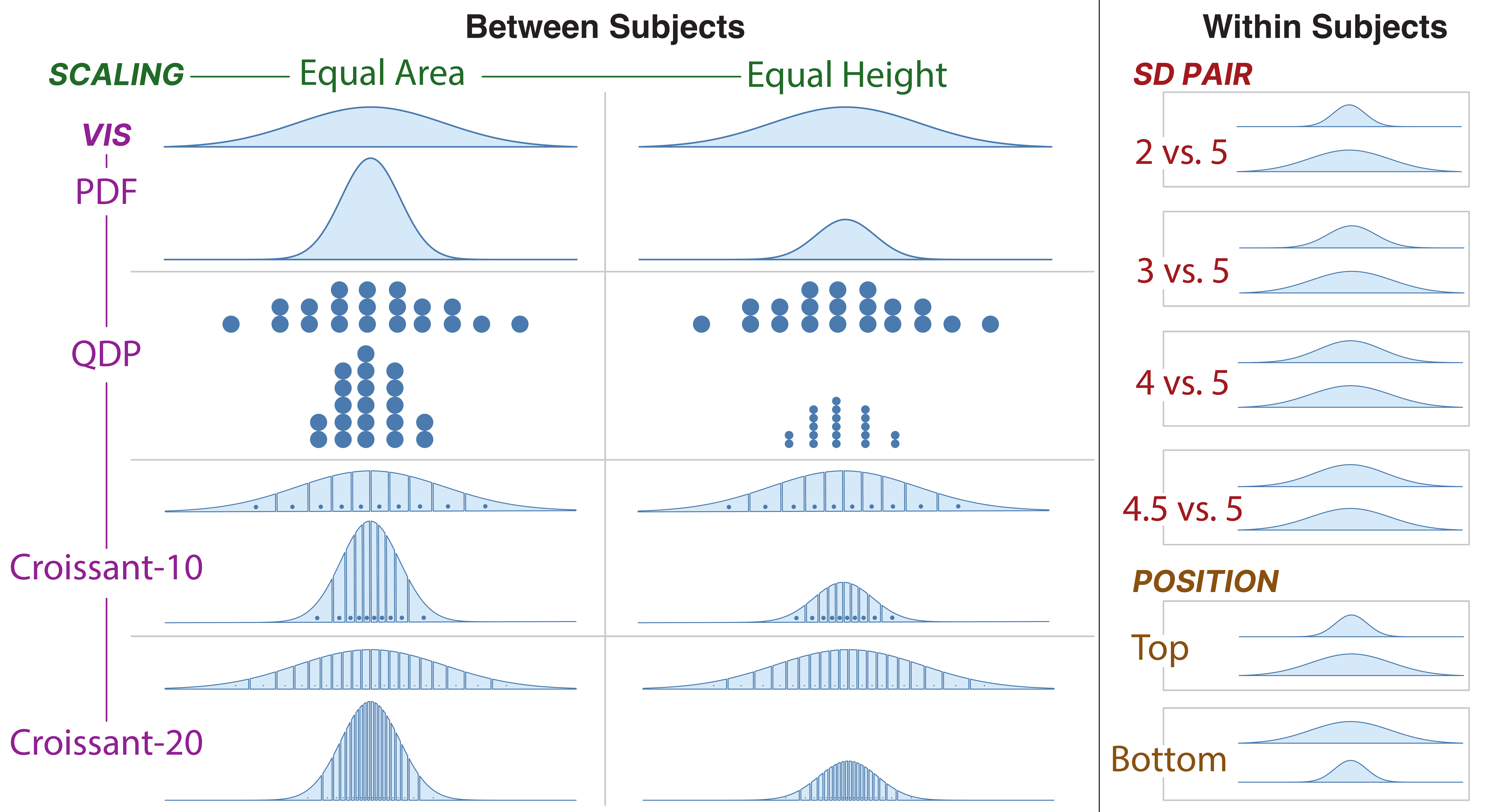}
\vspace{-2mm}
    \caption{
    Stimuli overview. Left: Between-subjects; different visualization techniques (PDF, QDP, Croissant-10, Croissant-20) under two scaling conditions (Equal Area, Equal Height). Right: within-subjects; standard deviation pairs and position variations.}
    \label{fig:stimuli}
    \vspace{-4mm}
\end{figure}

For all four graphs (PDFs, QDPs, Croissant-20s, and Croissant-10s), we created stimuli with two normal probability distributions stacked vertically (as shown in Fig.~\ref{fig:pdf-aff}, top). These stimuli were then scaled to either ``equal area'' or ``equal height'' (see Fig.~\ref{fig:stimuli}, left panel). For PDFs and both croissants ``equal area'' meant that the area below the two curves had the same number of pixels, and ``equal height'' meant that the distance from the highest point on the curve to the x-axis was held constant. For QDPs, ``equal area'' ensured that both graphs' dots were the same pixel size, while ``equal height'' maintained a constant distance from the top of the highest dot to the x-axis. In narrower distributions, this adjustment reduced dots' diameter to preserve their circular shape.

For each of our \revi{eight} stimuli types, we created four sets of distributions with varying SDs (see Fig.~\ref{fig:stimuli}, right panel). The first set presented a distribution with an SD of 2 and a distribution with an SD of 5. The rest compared distributions with SDs of 3 and 5, 4 and 5, and finally 4.5 and 5. We chose the first three of these SD comparisons because they were shown to have varying performance in previous work \cite{fygenson-padilla-pdf-scaling}. We added the 4.5 vs. 5 SD comparison, because we believed it could highlight QDPs' weaker affordance of a continuous distribution, as explained in H3 in Section~\ref{sec:hyp}. For each of the SD combinations we created two sets of stimuli, such that the vertical position of the narrower (higher SD) distribution was counterbalanced (see Fig.~\ref{fig:stimuli}, bottom right). This resulted in eight stimuli of each scaled visualization, for a total of 64 tested stimuli. We created all stimuli via custom D3.js functions, and edited (e.g., scaled, relabeled, etc.) them in Adobe Illustrator. Stimuli and generating scripts are available in Supplemental Materials.

\subsection{Experimental Design}
\label{sec:exp-design}
We designed a 4 (\textit{Visualization}) × 2 (\textit{Scaling}) × 4 (\textit{SD Pairs}) × 2 (\textit{Position}) mixed-subjects design \revi{(see Figure~\ref{fig:stimuli}).} The between-subjects variables were \textit{Visualization} and \textit{Scaling}. The within-subject variable of interest was \textit{SD Pairs}. We included \textit{Position} to counterbalance our stimuli and increase statistical power. We did not hypothesize about \textit{Position}'s impact, and included it as a covariate in our analysis. 
We investigated task correctness and collected participants' qualitative responses to observe afforded strategies.

\subsection{Participants}
We recruited U.S.-based participants over 18 through Prolific~\cite{prolific}, which anonymizes, qualifies, enrolls, and compensates participants, enabling a double-blind study. 
As per our preregistered power analysis, we recruited 808 participants (\textit{n} = 101 per between-subject group). Because affordances depend on readers' individual characteristics~\cite{fygenson-cog-affordance-framework}, we crowdsourced responses from people with varying backgrounds. 
To account for variance in visualization education, we collected participants' graph literacy scores using the Short Graph Literacy test~\cite{okan-sgl-2019}.

\subsection{Procedure}
\label{sec:procedure}
Participants were informed of the estimated duration and compensation (12 USD/hour) and consented to an IRB-approved form, after which they completed the study on Qualtrics~\cite{qualtrics}. First, participants read instructions and were informed of a scenario in which scientists are measuring the concentration of two different solutes in seawater. This scenario was identical to that used in previous PDF comparison work~\cite{fygenson-padilla-pdf-scaling}, and adapted from a study on HOPs~\cite{hullman-hops}. Then, participants \revi{were asked to answer} an initial attention check \revi{question to confirm they were reading the instructions. Next,} they were instructed to make the size of their browser window as large as possible and shown an annotated illustration of an example plot. Next, participants answered eight questions, one for each pair of charts. Following previously published procedures \cite{fygenson-padilla-pdf-scaling}, all the charts were presented in randomized order, and the questions read ``Which solute, if either, has a \textbf{higher probability} of being present at \textbf{X or less ppm} in the sampled seawater?'' where X was in the same position across all stimuli (at threshold \textit{t} in Figure~\ref{fig:pdf-aff}), but varied depending on the numbering of each stimulus's x-axis. Participants chose from three options (``solute A has a higher probability'', ``solute B has a higher probability,'' and ``neither solute A nor solute B has a higher probability''). The correct option was always the solute with a lower SD. This task does not represent the full range of probability distribution use cases, but allows us to investigate the extent to which affordances can impact strategies such that specific task performance is modulated.
After completing the chart comparison questions, participants used an open-ended text response to report their thought processes while reading the charts. Finally, participants completed Okan et al.'s Short Graph Literacy test~\cite{okan-sgl-2019}. 
The survey and its stimuli are available in Supplemental Materials.

\subsection{Analysis}
\label{sec:exp-analysis}
We preregistered the binomial Bayesian model in~\autoref{eq:model1} with uninformative priors centered at 0 and a standard deviation of 2.5. We used R packages \textit{tidyr} v. 1.3.1 for data processing \cite{r-tidyr}, \textit{brms} v. 2.20.4 for Bayesian modeling \cite{r-brms}, and \textit{tidybayes} v. 3.0.6 for data processing and visualization \cite{r-tidybayes}. \revi{As denoted in Eq.~\ref{eq:model1}}, our model 
assesses the variance in participants' binary correctness as explained by the interaction between \textit{Visualization} and \textit{Scaling}, and the interaction between \textit{Visualization} and \textit{SD Pair}, as well as their lower-order terms.  \revi{In this model, $p_i = P(Y_i = 1)$ represents the posterior probability that participant $i$ makes a correct comparison. Because we use a logit link, $\text{logit}(p_i) = \log\!\left(\frac{p_i}{1 - p_i}\right)$, all fixed-effect coefficients are estimated on the log-odds scale.} Variable levels are described in Sec.~\ref{sec:exp-design}. We also account for any variance explained by \textit{Position} and \textit{Graph Literacy}.
\revi{
\begin{equation}
\begin{aligned}
Y_i &\sim \text{Bernoulli}(p_i) \\
\text{logit}(p_i) &= 
\beta_0 + \beta_1 \text{Visualization}_i + \beta_2 \text{Scaling}_i \\
&\quad + \beta_3 (\text{Visualization}_i \times \text{Scaling}_i)
+ \beta_4 \text{SDPair}_i \\
&\quad + \beta_5 (\text{Visualization}_i \times \text{SDPair}_i)
+ \beta_6 \text{Position}_i \\
&\quad + \beta_7 \text{GraphLiteracy}_i
+ u_{ID[i]}
\end{aligned}
\label{eq:model1}
\end{equation}}
We also conducted an exploratory analysis of each visualization's affordances. Two raters qualitatively coded participants' descriptions of their thought processes when completing the cumulative probability task. We then calculated the intraclass correlation (ICC) for each code, and the raters reconciled any codes that had an ICC 95\% confidence interval with a lower bound less than 0.6. 

\begin{figure}[t!]
\centering  
\includegraphics[width=0.85\linewidth, alt={Visualized posteriors of the numbers in Table 1. The x-axis indicates likelihood of correct comparison, and range from 0 (left) to 100 (right).}]{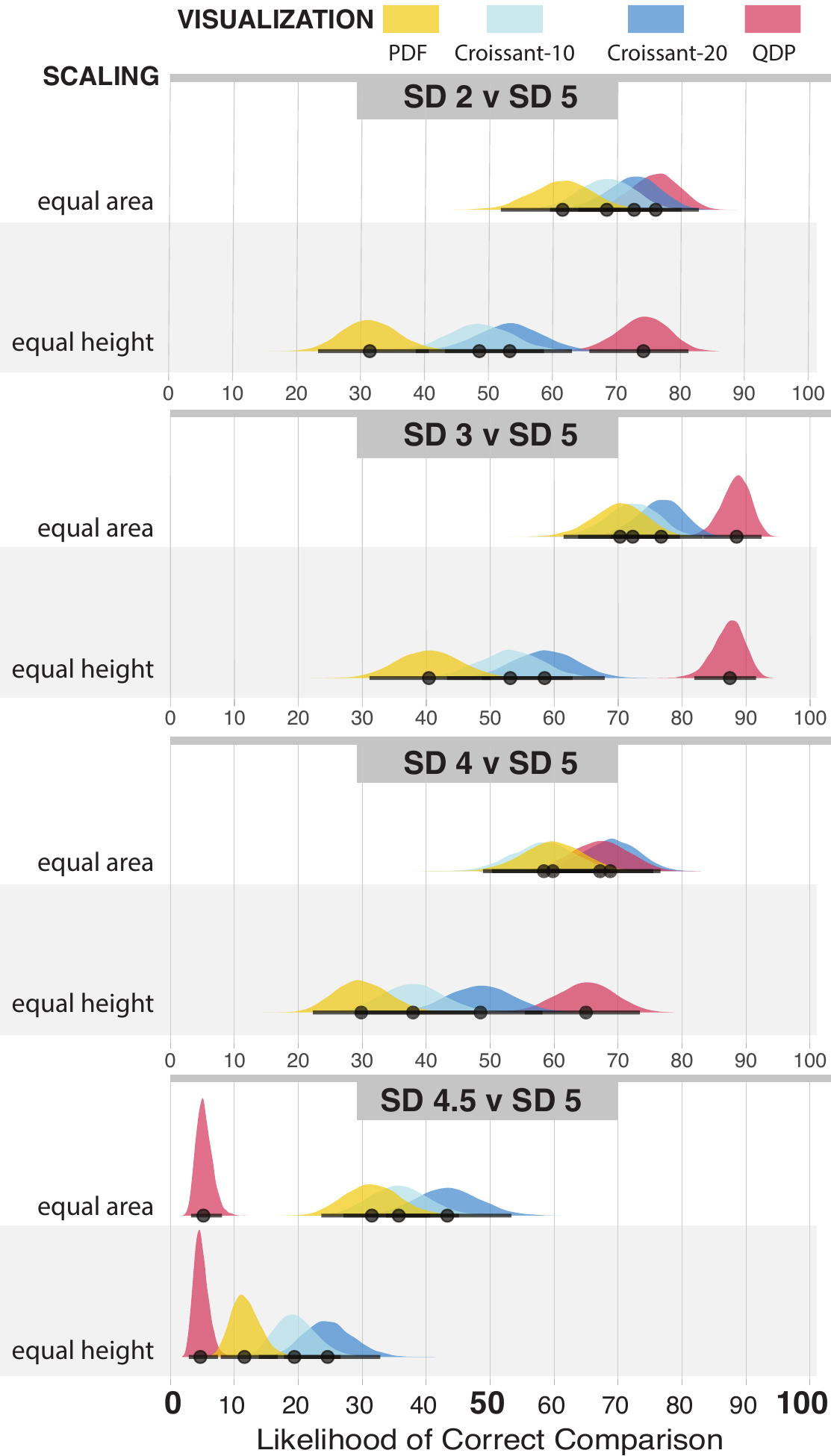}
    \caption{\revi{Posteriors of Eq.~\ref{eq:model1} show that QDPs are robust to equal-height scaling and perform best across most conditions except the 4.5 v 5 SD comparison, where they perform the worst. Equal-height croissant charts outperform equal-height PDFs. Distributions show posterior predicted probabilities of correct responses, transformed from log-odds 
    for interpretability. Black dots and lines show mean predicted probabilities 
    and 95\% credible intervals. 
    }}
    \label{fig:posteriors}
    \vspace{-9mm}
\end{figure}

\section{Experimental Results}
\label{sec:results}
Below, we describe the results from our analysis of binary task correctness, the results from our exploratory analysis of affordances, and the relationship between observed affordance and task correctness. As in our preregistered justification, we recruited 808 participants (\textit{n} = 101 for each between-subject group). 438 participants self-reported to not be men, with a mean age of 44.0 years (\textit{SD} = 15.4). The mean short graph literacy score was 2.3 of 4 (\textit{SD} = 1.1). \rev{Participants took an average of 11 minutes and 5 seconds (\textit{SD} = 7 min, 2 sec) to complete the survey.} 
As preregistered, we computed our model with and without participants who failed the attention check and performed a sensitivity analysis. There were no meaningful differences in the results between the two samples. For a large sample size and more conservative results, we report posteriors of the model that uses \revi{responses from participants who passed and those who failed the attention check}. 

\subsection{Task Performance: Investigating Hypotheses}
\label{sec:hyp-results}
We investigated the amount of variance in correct comparisons based on visualization, scaling, and SD pairs with the binomial model in Eq.~\ref{eq:model1}. As preregistered, we interpret interaction effects with credible intervals that exclude zero as accounting for meaningful variance in task correctness. We use the term ``meaningful'' instead of ``statistically significant'' to avoid implying a fixed decision threshold (e.g., \textit{p} < .05), as advised by Bayesian best practices \cite{kruschke2021bayesianreporting}. We ran our model with each combination of possible referents. 
All analyses are available in Supplemental Materials. 

To examine H1, if equal-height \pdf{PDFs} lead to meaningfully fewer correct comparisons than equal-area \pdf{PDFs}, we can look to \textit{Scaling}'s main effect when \textit{Visualization}'s referent is the \pdf{PDF} condition (row 1 in Table~\ref{tab:model1}). \textit{Scaling}'s credible interval does not include zero, indicating that scaling has a meaningful effect on \pdf{PDF} task performance. By visually examining the posteriors from this model (Fig.~\ref{fig:posteriors}), the results show that equal-height \pdf{PDFs} (colored yellow in rows with gray backgrounds) result in fewer correct answers than equal-area \pdf{PDFs} (yellow in rows with white backgrounds) across all SD Pairs. Thus, we accept H1, and present evidence for the reproducibility of results from Fygenson and Padilla \cite{fygenson-padilla-pdf-scaling}.

To examine H2a-d, how equal-height visualizations that afford counting impact task correctness in comparison to equal-height \pdf{PDFs}, we can refer to rows 5-16 in Table~\ref{tab:model1}. These rows show the main effects of \textit{Visualization} when the model is run with \pdf{PDF} as the referent. Equal-height \ctw{Croissant-20s} resulted in meaningfully improved task correctness in comparison to equal-height \pdf{PDFs} across all SD pairs (rows 9-12), leading us to accept H2b. Equal-height \qdp{QDPs} meaningfully improved task correctness compared to equal-height \pdf{PDFs} across all SD pairs (rows 5–7), except for the 4.5 vs. 5 pair, where \qdp{QDP} performance was meaningfully worse, as indicated by the sign change in the last three columns of row 8. These findings support the acceptance of H2c, contingent on the considerations outlined in H3. Equal-height \cten{Croissant-10s} resulted in meaningfully improved task correctness only over equal-height \pdf{PDFs} for SD pairs 2 vs. 5 and 4.5 vs. 5 (rows 13 and 16), leading us to partially accept hypothesis H2a. These results are also evident in Figure~\ref{fig:posteriors}'s gray-background rows, which show \pdf{PDFs} consistently trailing behind both \ctw{Croissant-20s} and \qdp{QDPs}. 

To directly investigate H2d--whether \qdp{QDPs}, \ctw{Croissant-20s}, and \cten{Croissant-10s} reduce performance differences between equal-area and equal-height scaling more effectively than \pdf{PDFs}--we can examine the interaction effects of \textit{Visualization} and \textit{Scaling} in rows 17-19 of Table~\ref{tab:model1}. The results indicate that \qdp{QDPs} meaningfully decrease the expected drop in performance associated with scaling from equal-area to equal-height in comparison to \pdf{PDFs} (row 17). However, \cten{Croissant-10s} and \ctw{Croissant-20s} show no meaningful difference in their interactions with \pdf{PDFs} (rows 18-19). We do not find evidence to support the hypothesis that \cten{Croissant-10s} and \ctw{-20s} can mitigate the performance drop associated with equal-height \pdf{PDFs}, and thus only partially accept H2d. 

\begin{figure}[b]
\centering  
\vspace{-5mm}
\includegraphics[width=0.8\linewidth, alt={Equal height Croissant-10s encourage readers to interpret partial areas when the threshold t falls inside a quantile, as opposed to cleanly between two quantiles. Equal height QDPs do not encourage the interpretation of partial areas when they threshold t falls between two columns of dots, which can lead to incorrect distribution comparisons.}]{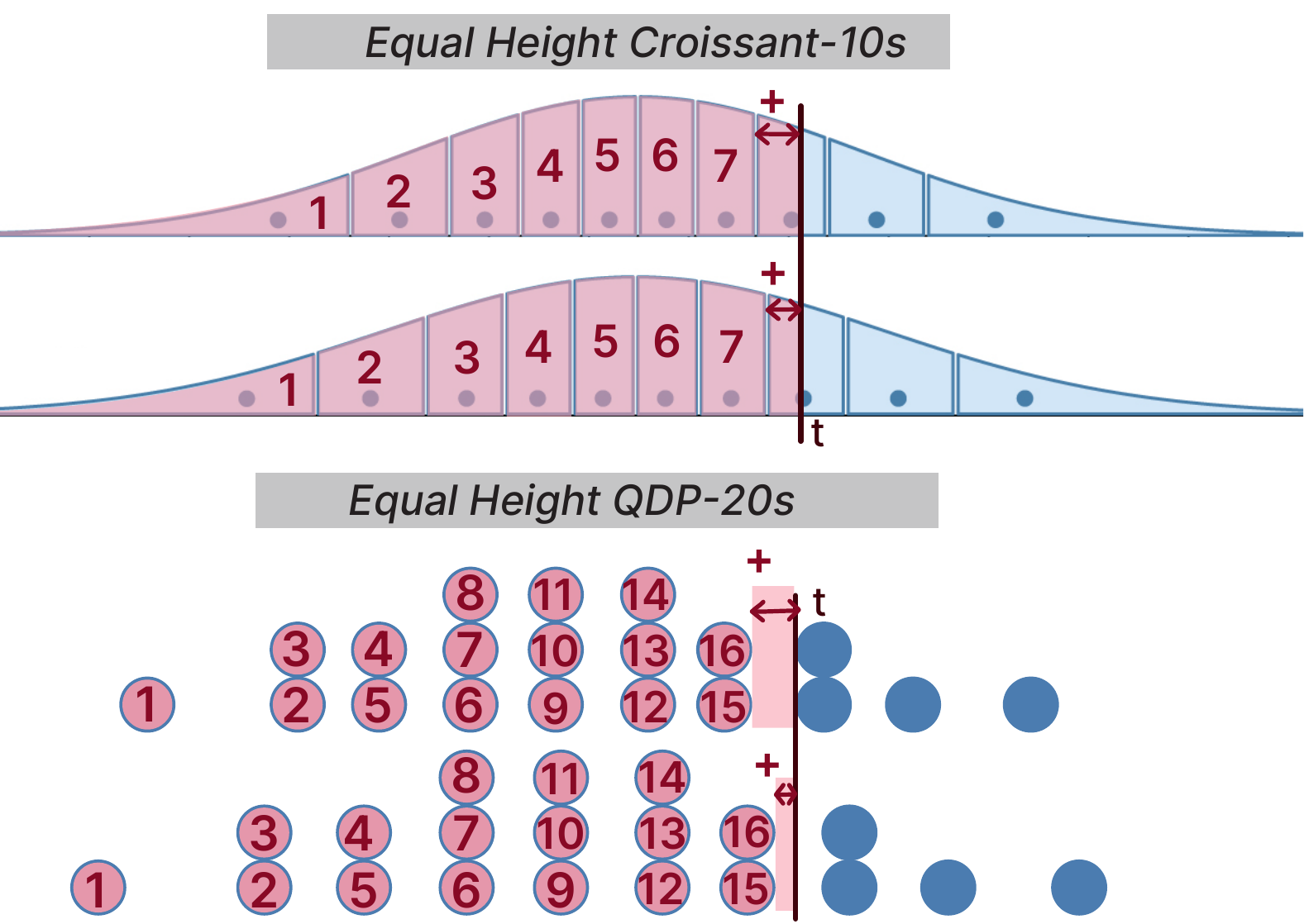}
\vspace{-2mm}
    \caption{\revi{Equal-height Croissant-10s and QDPs with an SD of 4.5 (tops) and an SD of 5 (bottoms).} Results show that Croissant-10s afford inter-edge interpolation much more than QDPs.}
    \label{fig:cut-off-dist}
    \vspace{-5mm}
\end{figure}

Lastly, to assess whether \qdp{QDPs} exhibit different task performance when comparing the 4.5 vs. 5 SD pair relative to other SD pairs, and if this performance differs from other visualizations
, we analyze the interaction between \textit{SD Pair} and \textit{Visualization}. Specifically, we examine meaningful differences when running the model with a \textit{Visualization} referent of \qdp{QDP} and \textit{SD Pair} referent of 4.5 vs. 5. Rows 20-28 in Table~\ref{tab:model1} show relevant model outputs and are all meaningfully different. A meaningful difference in QDP task performance across SD pairs is further supported by the last two rows of Figure~\ref{fig:posteriors}, where \qdp{QDPs} perform markedly worse than other charts. This drop in performance is notable because \qdp{QDPs} perform the best in all other conditions.  These findings lead us to accept (H3) that \qdp{QDPs}' relationship between task performance for 4.5 v 5 SD and any other SD pair is different from \pdf{PDFs}', \ctw{Croissant-20s}', and \cten{Croissant-10s}'. 
To further analyze QDPs' performance 
in the 4.5 v 5 \textit{SD Pair} condition, we examine the main effect of \textit{Visualization} when the model's referents are set to \qdp{QDP},  4.5 vs. 5 SD, and first Equal Height, then Equal Area. Rows 32-37 in Table~\ref{tab:model1} show that \qdp{QDPs} perform meaningfully worse than \pdf{PDFs}, \ctw{Croissant-20s}, and \cten{Croissant-10s} when comparing a distribution with SD of 4.5 to a distribution with SD of 5, regardless of scaling.

In summary, our analysis reveals that equal-height \pdf{PDFs} lead to a lower likelihood of task correctness than equal-area \pdf{PDFs}, equal-height \ctw{Croissant-20s} improved chances of correctness over equal-height \pdf{PDFs}, although not as well as equal-height \qdp{QDPs}. \qdp{QDPs} were especially robust to scaling, and frequently outperformed other visualizations. However, \qdp{QDPs} led to meaningfully lower chances of correctly comparing distributions with SDs of 4.5 and 5, than all other visualizations. These findings highlight the benefits of \ctw{Croissant-20s} and \qdp{QDPs} over \pdf{PDFs}, while also demonstrating a specific weakness of low-quantile \qdp{QDPs}.
\vspace{-2mm}

\begin{table}[h!]
 \caption{
  \revi{Rows report posterior log-odds from Eq.~\protect{\ref{eq:model1}} contrasted between experimental conditions with 95\% Bayesian credible intervals. Darker cell backgrounds indicate stronger effects. l-CI and u-CI denote lower and upper bounds of credible intervals, respectively. Cr-20 = Croissant-20; Cr-10 = Croissant-10. Odds ratios are obtained by exponentiating the reported log-odds estimates.}
  }
    \vspace{-1mm}
    {\footnotesize
    \resizebox{\columnwidth}{!}{
    \begin{tabular}{@{}l@{\hspace{0.05cm}}c@{\hspace{0.15cm}}c@{\hspace{0.18cm}}c| l@{\hspace{0.2cm}}c@{\hspace{0.22cm}}c@{\hspace{0.15cm}}c@{}}
        \toprule
        \multicolumn{4}{c|}{Referent Conditions} & Comparison & Est\rev{imate} & \textbf{l-CI} & \textbf{u-CI} \\ 
        \midrule
         \textit{Row} & \textbf{Vis} & \textbf{SD Pair} & \textbf{Scaling} & & & & \\
        \midrule
        \textit{1} & PDF & N/A & Eq. Height & Eq. Area & \cellcolor[HTML]{DBE7F6}{1.26} & \cellcolor[HTML]{DBE7F6}0.78 & \colorbox[HTML]{DBE7F6}{1.75}\\
        \textit{2} & QDP & N/A & Eq. Height & Eq. Area & 0.11 & -0.37 & 0.61 \\
        \textit{3} & Cr-20 & N/A & Eq. Height & Eq. Area & \cellcolor[HTML]{EDF3FB}0.86 & \cellcolor[HTML]{EDF3FB}0.38 & \colorbox[HTML]{EDF3FB}{1.34}\\
        \textit{4} & Cr-10 & N/A & Eq. Height & Eq. Area & \cellcolor[HTML]{EDF3FB}{0.82} & \cellcolor[HTML]{EDF3FB}0.35 & \colorbox[HTML]{EDF3FB}{1.30}\\
        \midrule
        \textit{5} & PDF & 2 v 5 & Eq. Height & QDP & \cellcolor[HTML]{DBE7F6}1.80 & \cellcolor[HTML]{DBE7F6}1.24 & \colorbox[HTML]{DBE7F6}{2.37} \\
        \textit{6} & PDF & 3 v 5 & Eq. Height & QDP & \cellcolor[HTML]{C4DAF1}2.29 & \cellcolor[HTML]{C4DAF1}1.71 & \colorbox[HTML]{C4DAF1}{2.88} \\
        \textit{7} & PDF & 4 v 5 & Eq. Height & QDP & \cellcolor[HTML]{DBE7F6}1.45 & \cellcolor[HTML]{DBE7F6}0.88 & \colorbox[HTML]{DBE7F6}{2.02}\\
        \textit{8} & PDF & 4.5 v 5 & Eq. Height & QDP & \cellcolor[HTML]{EDF3FB}-0.99 & \cellcolor[HTML]{EDF3FB}-1.65 & \colorbox[HTML]{EDF3FB}{-0.34} \\
        \textit{9} & PDF & 2 v 5 & Eq. Height & Cr-20 & \cellcolor[HTML]{EDF3FB}0.88 & \cellcolor[HTML]{EDF3FB}0.32 & \colorbox[HTML]{EDF3FB}{1.45} \\
        \textit{10} & PDF & 3 v 5 & Eq. Height & Cr-20 & \cellcolor[HTML]{EDF3FB}0.71 & \cellcolor[HTML]{EDF3FB}0.15 & \colorbox[HTML]{EDF3FB}{1.28}\\
        \textit{11} & PDF & 4 v 5 & Eq. Height & Cr-20 & \cellcolor[HTML]{EDF3FB}0.77 & \cellcolor[HTML]{EDF3FB}0.21 & \colorbox[HTML]{EDF3FB}{1.33}\\
        \textit{12} & PDF & 4.5 v 5 & Eq. Height & Cr-20 & \cellcolor[HTML]{EDF3FB}0.90 & \cellcolor[HTML]{EDF3FB}0.32 & \colorbox[HTML]{EDF3FB}{1.49}\\
        \textit{13} & PDF & 2 v 5 & Eq. Height & Cr-10 & \cellcolor[HTML]{EDF3FB}0.69 & \cellcolor[HTML]{EDF3FB}0.13 & \colorbox[HTML]{EDF3FB}{1.25} \\
        \textit{14} & PDF & 3 v 5 & Eq. Height & Cr-10 & 0.49 & -0.06 & 1.05 \\
        \textit{15} & PDF & 4 v 5 & Eq. Height & Cr-10 & 0.34 & -0.22 & 0.89 \\
        \textit{16} & PDF & 4.5 v 5 & Eq. Height & Cr-10 & \cellcolor[HTML]{EDF3FB}0.60 & \cellcolor[HTML]{EDF3FB}0.02 & \colorbox[HTML]{EDF3FB}{1.19} \\
        \midrule
        \textit{17} & PDF & N/A & Eq. Height & QDP x Eq. Area & \cellcolor[HTML]{DBE7F6}-1.14 & \cellcolor[HTML]{DBE7F6}-1.83 & \colorbox[HTML]{DBE7F6}{-0.46} \\
        \textit{18} & PDF & N/A & Eq. Height & Cr-20 x Eq. Area & -0.40 & -1.09 & 0.27 \\
        \textit{19} & PDF & N/A & Eq. Height & Cr-10 x Eq. Area & -0.40 & -1.08 & 0.27 \\
        \midrule
        \textit{20} & QDP & 4.5 v 5 & N/A & PDF x 2 v 5 & \cellcolor[HTML]{C4DAF1}-2.61 & \cellcolor[HTML]{C4DAF1}-3.17 & \colorbox[HTML]{C4DAF1}{-2.05}\\
        \textit{21} & QDP & 4.5 v 5 & N/A & PDF x 3 v 5 & \cellcolor[HTML]{A5C8EA}-3.08 & \cellcolor[HTML]{A5C8EA}-3.66 & \colorbox[HTML]{A5C8EA}{-2.49}\\
        \textit{22} & QDP & 4.5 v 5 & N/A & PDF x 4 v 5 & \cellcolor[HTML]{C4DAF1}-2.24 & \cellcolor[HTML]{C4DAF1}-2.80 & \colorbox[HTML]{C4DAF1}{-1.69}\\
        \textit{23} & QDP & 4.5 v 5 & N/A & Cr-20 x 2 v 5 & \cellcolor[HTML]{C4DAF1}-2.58 & \cellcolor[HTML]{C4DAF1}-3.14 & \colorbox[HTML]{C4DAF1}{-2.04}\\
        \textit{24} & QDP & 4.5 v 5 & N/A & Cr-20 x 3 v 5 & \cellcolor[HTML]{A5C8EA}-3.24 & \cellcolor[HTML]{A5C8EA}-3.82 & \colorbox[HTML]{A5C8EA}{-2.66}\\
        \textit{25} & QDP & 4.5 v 5 & N/A & Cr-20 x 4 v 5 & \cellcolor[HTML]{C4DAF1}-2.35 & \cellcolor[HTML]{C4DAF1}-2.89 & \colorbox[HTML]{C4DAF1}{-1.79}\\
        \textit{26} & QDP & 4.5 v 5 & N/A & Cr-10 x 2 v 5 & \cellcolor[HTML]{C4DAF1}-2.47 & \cellcolor[HTML]{C4DAF1}-3.03 & \colorbox[HTML]{C4DAF1}{-1.92}\\
        \textit{27} & QDP & 4.5 v 5 & N/A & Cr-10 x 3 v 5 & \cellcolor[HTML]{A5C8EA}-3.15 & \cellcolor[HTML]{A5C8EA}-3.74 & \colorbox[HTML]{A5C8EA}{-2.58}\\
        \textit{28} & QDP & 4.5 v 5 & N/A & Cr-10 x 4 v 5 & \cellcolor[HTML]{C4DAF1}-2.48 & \cellcolor[HTML]{C4DAF1}-3.02 & \colorbox[HTML]{C4DAF1}{-1.94}\\
        \midrule
        \textit{29} & QDP & 4.5 v 5 & N/A & 2 v 5 & \cellcolor[HTML]{A5C8EA}3.86 & \cellcolor[HTML]{A5C8EA}3.43 & \colorbox[HTML]{A5C8EA}{4.31}\\
        \textit{30} & QDP & 4.5 v 5 & N/A & 3 v 5 & \cellcolor[HTML]{7BAFDB}4.73 & \cellcolor[HTML]{7BAFDB}4.27 & \colorbox[HTML]{7BAFDB}{5.21}\\
        \textit{31} & QDP & 4.5 v 5 & N/A & 4 v 5 & \cellcolor[HTML]{A5C8EA}3.43 & \cellcolor[HTML]{A5C8EA}3.02 & \colorbox[HTML]{A5C8EA}{3.87} \\
        \midrule
        \textit{32} & QDP & 4.5 v 5 & Eq. Area & PDF & \cellcolor[HTML]{DBE7F6}1.92 & \cellcolor[HTML]{DBE7F6}1.32 & \colorbox[HTML]{DBE7F6}{2.53}\\
        \textit{33} & QDP & 4.5 v 5 & Eq. Area & Cr-20 &\cellcolor[HTML]{C4DAF1} 2.39 & \cellcolor[HTML]{C4DAF1}1.79 & \colorbox[HTML]{C4DAF1}{3.00}\\
        \textit{34} & QDP & 4.5 v 5 & Eq. Area & Cr-10 & \cellcolor[HTML]{C4DAF1}2.07 & \cellcolor[HTML]{C4DAF1}1.47 & \colorbox[HTML]{C4DAF1}{2.68}\\
        \textit{35} & QDP & 4.5 v 5 & Eq. Height & PDF & \cellcolor[HTML]{EDF3FB}0.78 & \cellcolor[HTML]{EDF3FB}0.15 & \colorbox[HTML]{EDF3FB}{1.40}\\
        \textit{36} & QDP & 4.5 v 5 & Eq. Height & Cr-20 & \cellcolor[HTML]{DBE7F6}1.69 & \cellcolor[HTML]{DBE7F6}1.09 & \colorbox[HTML]{DBE7F6}{2.31}\\
        \textit{37} & QDP & 4.5 v 5 & Eq. Height & Cr-10 & \cellcolor[HTML]{DBE7F6}1.39 & \cellcolor[HTML]{DBE7F6}0.78 & \colorbox[HTML]{DBE7F6}{2.00}\\
        \bottomrule
    \end{tabular}
    }
    }
\label{tab:model1}
\vspace{-5mm}
\end{table}

\subsection{Affordances Described in Open-Ended Responses}
\label{sec:aff-results}
Two raters qualitatively coded strategies from participants' open-ended responses. The raters also identified responses likely AI-generated when they referenced elements not present in the stimuli (e.g., labels, y-axes). Of the 808 responses, \revi{21 were identified by both and excluded from coding. We include the codes of all other responses for a more conservative analysis. All qualitative coding data and analyses are available in the Supplemental Materials.}

\begin{figure}[b!]
\centering  
\vspace{-2mm}
\includegraphics[width=0.9\linewidth, alt={An annotated heat map shows the percent of participants that saw each of the visualization conditions (QDP, PDF, Croissant-20, and Croissant-10) and reported using different strategies when comparing distributions (counting, comparing height, area, spread, inter-edge interpolation, and whether the visual objects were directly over a a tick mark). QDPs show a high concentration (68\%) of participants reporting a counting strategy, whereas PDFs show no counting strategies, and both croissants led to ~20 to 30\% of participants reporting counting.  Additionally, 10\% of  croissant-10 participants reported using inter-edge interpolation, whereas only 0 to 1\% of other conditions' participants reported the same. PDFs and both croissants show a range of strategies reported, whereas QDPs show only one strategy (counting) reported by 68\% of participants and other strategies reported by between 1 and 4\% of participants. The last column (1 to 4\%) shows percentage of participant responses we believe to be AI-written, across visualization condition.}]{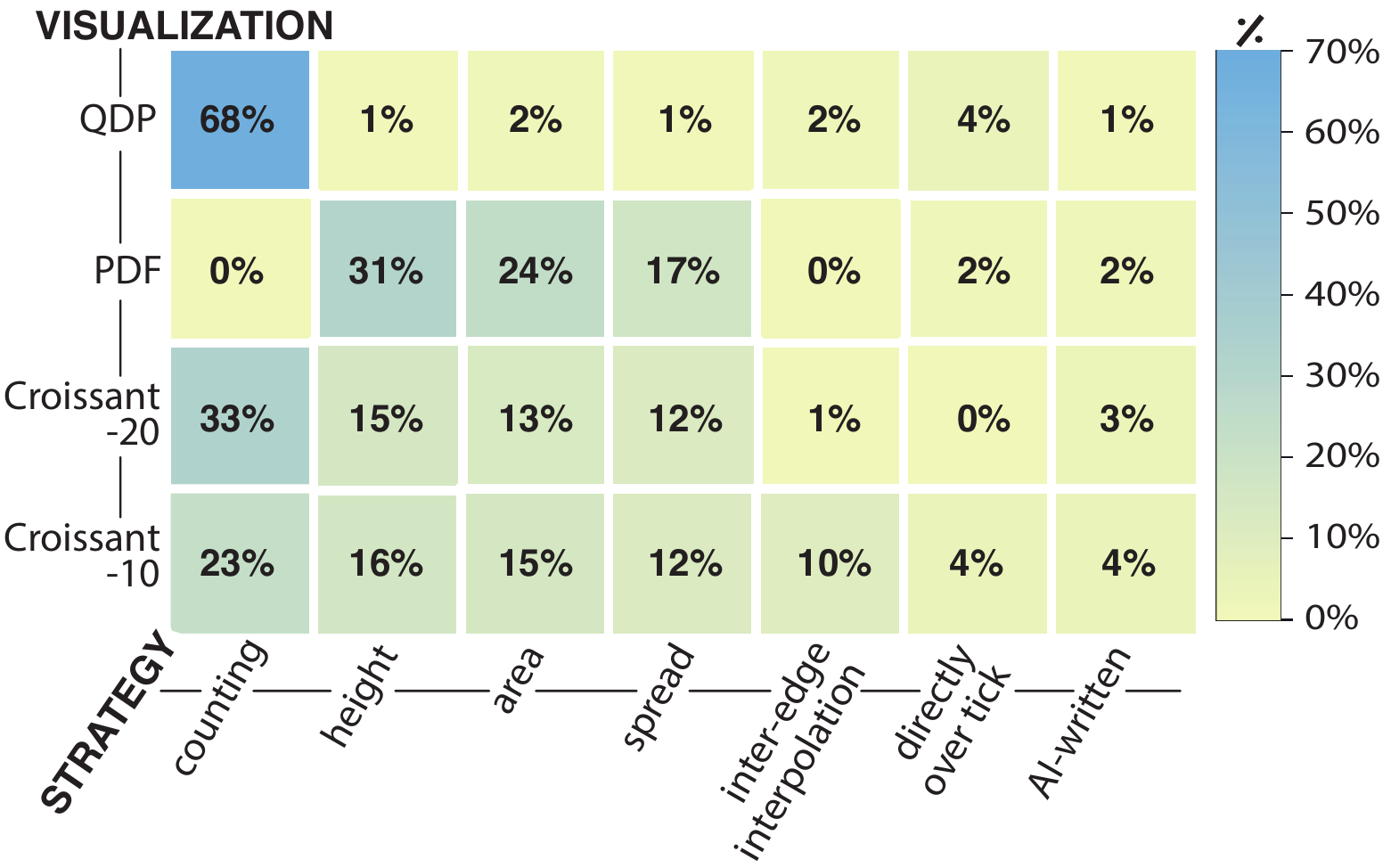}
\vspace{-2mm}
    \caption{Percent of each visualization that led to each strategy (columns) across tested visualizations (rows). 
    \rev{Strategies are not mutually exclusive so rows and columns may not equal 100\%.}}
    \label{fig:aff-counts}
    \vspace{-5mm}
\end{figure}

The \revi{review of open-ended responses} identified six strategies (mean ICC = 0.84, SD = 0.08): counting, height comparison, area comparison, spread comparison, judgment of how far into a visual mark a threshold occurred (i.e., ``inter-edge interpolation'', Fig.~\ref{fig:cut-off-dist}), and comparison of only marks directly above the threshold (i.e., ``directly over tick''). We show the percentage of participants who mentioned each strategy by \textit{Visualization} condition in Figure~\ref{fig:aff-counts}.

The results indicate that \qdp{QDPs} afforded counting (68\% of participants) more than any other visualization. This 
affordance aligns with initial pilot data and \qdp{QDPs}' frequency-framing motivation~\cite{kay-qdps}. \cten{Croissant-10s}~(23\%) and \ctw{-20s}~(33\%) resulted in roughly \revi{one-third to one-fourth} the number of counting strategies as \qdp{QDPs}, indicating a present, but weaker affordance. \ctw{C}\cten{r}\ctw{o}\cten{i}\ctw{s}\cten{s}\ctw{a}\cten{n}\ctw{t} \cten{c}\ctw{h}\cten{a}\ctw{r}\cten{t}\ctw{s}' counting affordance was exhibited in responses, such as ``... I looked at the number of 5\% `blocks' [that] were on one side on both charts.'' 
We did not observe any evidence that \pdf{PDFs} afforded counting.
 Figure~\ref{fig:aff-counts} also shows that there was very little evidence that \qdp{QDPs} afforded multiple strategies, in contrast to the other three visualizations.
This pattern indicates that \ctw{c}\cten{r}\ctw{o}\cten{i}\ctw{s}\cten{s}\ctw{a}\cten{n}\ctw{t} \cten{c}\ctw{h}\cten{a}\ctw{r}\cten{t}\ctw{s} and \pdf{PDFs} result in greater flexibility in strategy use than \qdp{QDPs}.

\subsection{Task Performance and Observed Affordances}
\label{sec:task-aff-correlation}
The results in Figure~\ref{fig:aff-counts} show that \qdp{QDPs} strongly afforded counting strategies, 
\ctw{Croissant-20s} and \cten{-10s} weakly afforded counting, and \pdf{PDFs} did not afford counting at all. When looking at the posteriors of Equation~\ref{eq:model1} in Figure~\ref{fig:posteriors}, we can see that the difference in the strength of visualizations' counting affordances corresponds with their likelihood of leading to correct cumulative probability comparisons between three of the four equal-height conditions we tested. These conditions are 2 vs. 5, 3 vs. 5, and 4 vs. 5 SD pairs, and their relative performance is remarkably consistent across equal-height conditions, which indicates that affording counting strategies could improve cumulative comparisons. 

Additionally, our quantitative results show that \qdp{QDPs} lead to the lowest likelihood of correct comparisons of distributions with SDs of 4.5 and 5. At the same time, our results exhibit that \qdp{QDPs} lack diversity in their afforded \revi{mental actions}. \pdf{PDFs}, \ctw{Croissant-20s} and \cten{-10s}, on the other hand, afford a variety of strategies. For the 4.5 vs. 5 SD Pair condition, our experimental findings illustrate that visualizations that afford noncounting strategies lead to a higher chance of correct comparisons.

Lastly, equal-height \pdf{PDFs} perform meaningfully worse than equal-area \pdf{PDFs} across all SD Pair conditions, and \pdf{PDFs}' most popular strategy is height comparison. To reason about this decrease, we can refer to the difference in height comparisons' efficacy across equal-area and equal-height \pdf{PDFs}.
Although comparing the peak height of two equal-area \pdf{PDFs} leads to consistently correct judgments of our tested task, peak height comparison between equal-height \pdf{PDFs} does not (Fig.~\ref{fig:pdf-aff}, row a).

\section{Discussion and Limitations}
\label{sec:discussion}
We present an affordance-motivated analysis of PDFs, followed by empirical evidence of the performance and affordances of area-encoded probability distribution visualizations. We do so through a case study of PDFs, which restricts the scope of our findings in the interest of examining the relationship between performance and affordance. Below, we detail the concrete contributions of this work, posit potential extrapolations of our findings, and discuss the limitations of our scoping and methodology, before describing future avenues of affordance research in visualization.

\subsection{Contributions to Affordance Research in Visualizations}
We analyze the relationship between performance and affordance evidence from our study to offer insights about the drivers behind previously documented comparison errors in PDFs \cite{fygenson-padilla-pdf-scaling} and performance improvements from QDPs \cite{kay-qdps, fernandes2018-qdp-cdf, kale2020-qdp}. We found a strong correlation between counting and equal-height visualizations' performance, indicating that counting supports cumulative probability comparisons.
We believe that \qdp{QDPs}' and \ctw{c}\cten{r}\ctw{o}\cten{i}\ctw{s}\cten{s}\ctw{a}\cten{n}\ctw{t} \cten{c}\ctw{h}\cten{a}\ctw{r}\cten{t}\ctw{s}' counting affordance provides indirect evidence of frequency-framing at work; 
however, we caution that this conclusion is circumstantial, as participants often struggle to fully articulate their cognitive processes\cite{Nisbett1977-NISTMT}.

Another example of affordance-performance alignment is seen in \cten{Croissant-10s} outperforming \qdp{QDPs} in the 4.5 vs. 5 SD condition, and \cten{Croissant-10s}' stronger affordance of ``inter-edge interpolation.'' This \revi{term} refers to participants counting both the whole units prior to a threshold, and the partial amount of a visual mark that is bisected by a threshold (see Fig.~\ref{fig:cut-off-dist}). Inter-edge interpolation suggests that participants recognize that probability does not change in discrete steps along the boundaries of visual marks, but instead forms a continuous distribution. One of our goals \revi{in designing} croissant charts was to afford inter-edge interpolation, because we hypothesized that \qdp{QDPs} weakly afford this interpolation, which can lead to errors when comparing similar SD pairs.  

We also see that \ctw{Croissant-20s} outperform \qdp{QDPs} in the 4.5 vs. 5 SD condition. This may be because \ctw{Croissant-20s} show higher resolution along the x-axis than the \qdp{QDPs} we tested. Although both charts encode 20 quantiles (i.e., each visual mark equals 5\% probability), \qdp{QDPs} stack quantiles such that moving from one column to the next can jump the cumulative probability by more than one quantile, affording discrete changes that may not exist. When designing \ctw{c}\cten{r}\ctw{o}\cten{i}\ctw{s}\cten{s}\ctw{a}\cten{n}\ctw{t} \cten{c}\ctw{h}\cten{a}\ctw{r}\cten{t}\ctw{s}, we sliced them into vertical quantiles, as opposed to \qdp{QDPs}' vertical and horizontal binning technique. 
A benefit of this design is afforded continuity and increased detail along the x-axis such that close SD comparisons are more supported. On the other hand, \qdp{QDPs}' binning technique affords counting much more strongly than \ctw{c}\cten{r}\ctw{o}\cten{i}\ctw{s}\cten{s}\ctw{a}\cten{n}\ctw{t}\cten{s}', leading to \qdp{QDPs}' superior performance across less similar SD Pairs. Additionally, increasing \qdp{QDPs}' quantiles would increase their resolution, possibly improving their performance in the 4.5 vs. 5 SD condition.

\subsection{Limitations and Future Work}
\label{sec:limitations}
We present a case study of probability visualizations that express distribution shape, which is inherently limited in scope. 
Additionally, we only investigate normal \revi{static} distributions because our case study is built on previous findings with normal distributions~\cite{fygenson-padilla-pdf-scaling}, and because non-normal distributions are significantly more difficult 
to interpret~\cite{peterson1964mode}. Future work should examine skewed and multimodal visualizations to increase the usefulness 
of this research. We also restrict our experiment to investigate a single, common task. Investigating other tasks, such as readers' accurate estimation of cumulative probabilities, 
and decision-making, would expand understanding of how a single affordance may correlate to a range of tasks.

Additionally, we used open-ended responses to uncover afforded strategies, which restricts our affordance findings to strategies that participants are conscious of and able to report. Research in psychology cautions against soley using participant-reported evidence, because of their lack of correlation to participant decision-making \cite{Nisbett1977-NISTMT}. Still, this open-ended reporting is useful for generating hypotheses and corroborating the confirmatory behavioral findings. 

Lastly, expanding the study of the relationship between affordance and performance to other types of visualization and other kinds of information is warranted. This expansion would not only broaden our understanding of other visualizations and how their designs drive reader conclusions, but can also be used to hypothesize about visual techniques that can be adopted for probability communication. For example, spatial subdivision in pie charts and tree maps was instrumental to our affordance-motivated PDF redesign.

\section{Conclusion}
We present a case study of cognitive affordances and performance in 
probability distribution visualizations and demonstrate a relationship between the two metrics. In turn, our findings suggest that cognitive affordances not only exist in visualizations, but also can inform hypotheses about their task performance. This work contributes to the growing body of affordance work in visualization, and establishes empirical evidence for the utility of affordances in reasoning about visualization performance and \revi{mental actions}.

\section{Acknowledgments}
This work was funded in part by NSF Grant \#2428149.

\bibliographystyle{eg-alpha-doi} 
\bibliography{egbibsample}

@article{garcia2017designing,
	address = {University of Granada, Granada, Spain.; University of Oklahoma, Norman.},
	author = {Garcia-Retamero, Rocio and Cokely, Edward T},
	crdt = {2017/02/14 06:00},
	date = {2017 Jun},
	dcom = {20180312},
	dep = {20170213},
	doi = {10.1177/0018720817690634},
	edat = {2017/02/14 06:00},
	issn = {1547-8181 (Electronic); 0018-7208 (Linking)},
	jid = {0374660},
	journal = {Hum Factors},
	jt = {Human factors},
	lid = {10.1177/0018720817690634 {$[$}doi{$]$}},
	lr = {20231031},
	mhda = {2018/03/13 06:00},
	month = {Jun},
	number = {4},
	oto = {NOTNLM},
	own = {NLM},
	pages = {582--627},
	phst = {2017/02/14 06:00 {$[$}pubmed{$]$}; 2018/03/13 06:00 {$[$}medline{$]$}; 2017/02/14 06:00 {$[$}entrez{$]$}},
	pl = {United States},
	pmid = {28192674},
	pst = {ppublish},
	pt = {Journal Article; Research Support, Non-U.S. Gov't; Research Support, U.S. Gov't, Non-P.H.S.; Systematic Review},
	sb = {IM},
	status = {MEDLINE},
	title = {Designing Visual Aids That Promote Risk Literacy: A Systematic Review of Health Research and Evidence-Based Design Heuristics.},
	volume = {59},
	year = {2017}}

@article{ancker2006design,
  title={Design features of graphs in health risk communication: a systematic review},
  author={Ancker, Jessica S and Senathirajah, Yalini and Kukafka, Rita and Starren, Justin B},
  journal={Journal of the American Medical Informatics Assoc.},
  volume={13},
  number={6},
  pages={608--618},
  year={2006},
  publisher={BMJ Group BMA House, Tavistock Square, London, WC1H 9JR}
}

@article{kong2010perceptual,
author = {Kong, Nicholas and Heer, Jeffrey and Agrawala, Maneesh},
title = {Perceptual Guidelines for Creating Rectangular Treemaps},
year = {2010},
publisher = {IEEE Educational Activities Department},
address = {USA},
volume = {16},
number = {6},
issn = {1077-2626},
url = {https://doi.org/10.1109/TVCG.2010.186},
doi = {10.1109/TVCG.2010.186},
journal = {IEEE Transactions on Visualization and Computer Graphics},
pages = {990–998},
numpages = {9}
}

@article{yang2023swaying,
  author={Yang, Fumeng and Cai, Mandi and Mortenson, Chloe and Fakhari, Hoda and Lokmanoglu, Ayse D. and Hullman, Jessica and Franconeri, Steven and Diakopoulos, Nicholas and Nisbet, Erik C. and Kay, Matthew},
  journal={IEEE Transactions on Visualization and Computer Graphics}, 
  title={Swaying the Public? {I}mpacts of Election Forecast Visualizations on Emotion, Trust, and Intention in the 2022 {U.S.} Midterms}, 
  year={2024},
  volume={30},
  number={1},
  pages={23-33}}

@inbook{padilla_review2022,
author = {Padilla, Lace and Kay, Matthew and Hullman, Jessica},
publisher = {Wiley StatsRef: Statistics Reference Online},
isbn = {9781118445112},
title = {Uncertainty Visualization},
booktitle = {Wiley StatsRef: Statistics Reference Online},
chapter = {},
pages = {1-18},
doi = {10.1002/9781118445112.stat08296},
year = {2022}
}

@article{mackinlay-auto-design,
author = {Mackinlay, Jock},
title = {Automating the design of graphical presentations of relational information},
year = {1986},
issue_date = {April 1986},
publisher = {Assoc. for Computing Machinery},
address = {New York, NY, USA},
volume = {5},
number = {2},
issn = {0730-0301},
doi = {10.1145/22949.22950},
journal = {ACM Transactions on Graphics},
month = {apr},
pages = {110–141},
numpages = {32}
}

@ARTICLE{besher-feiner-autovisual,
  author={Beshers, C. and Feiner, S.},
  journal={IEEE Computer Graphics and Applications}, 
  title={AutoVisual: rule-based design of interactive multivariate visualizations}, 
  year={1993},
  volume={13},
  number={4},
  pages={41-49},
  doi={10.1109/38.219450}}

@article{padilla2022know,
author = {Padilla, Lace},
title = {Know your experimental uncertainty},
year = {2022},
issue_date = {November - December 2022},
publisher = {Assoc. for Computing Machinery},
address = {New York, NY, USA},
volume = {29},
number = {6},
issn = {1072-5520},
doi = {10.1145/3564022},
journal = {Interactions},
month = {nov},
pages = {21–23},
numpages = {3}
}

@ARTICLE{joslyn2021,
AUTHOR={Joslyn, Susan and Savelli, Sonia},   
TITLE={Visualizing Uncertainty for Non-Expert End Users: The Challenge of the Deterministic Construal Error},      	
JOURNAL={Frontiers in Computer Science},      
VOLUME={2},      
PAGES={58},     
YEAR={2021},      
DOI={10.3389/fcomp.2020.590232},      
ISSN={2624-9898}
}

@article{correll2014error,
  title={Error bars considered harmful: Exploring alternate encodings for mean and error},
  author={Correll, Michael and Gleicher, Michael},
  journal={IEEE Transactions on Visualization and Computer Graphics},
  volume={20},
  number={12},
  pages={2142--2151},
  year={2014},
  publisher={IEEE},
doi={10.1109/TVCG.2014.2346298}
}

@article{van2019communicating,
  title={Communicating uncertainty about facts, numbers and science},
  author={Van Der Bles, Anne Marthe and Van Der Linden, Sander and Freeman, Alexandra LJ and Mitchell, James and Galvao, Ana B and Zaval, Lisa and Spiegelhalter, David J},
  journal={Royal Society Open Science},
  volume={6},
  number={5},
  pages={181870},
  year={2019},
  publisher={The Royal Society},
doi={10.1098/rsos.181870}
}

@article{peterson1964mode,
  title={Mode, median, and mean as optimal strategies.},
  author={Peterson, Cameron and Miller, Alan},
  journal={Journal of Experimental Psychology},
  volume={68},
  number={4},
  pages={363},
  year={1964},
  publisher={American Psychological Assoc.},
doi={10.1037/h0040387}
}

@article{kale2020-qdp,
  author={Kale, Alex and Kay, Matthew and Hullman, Jessica},
  journal={IEEE Transactions on Visualization and Computer Graphics}, 
  title={Visual Reasoning Strategies for Effect Size Judgments and Decisions}, 
  year={2021},
  volume={27},
  number={2},
  pages={272-282},
  doi={10.1109/TVCG.2020.3030335}}

@inproceedings{fernandes2018-qdp-cdf,
author = {Fernandes, Michael and Walls, Logan and Munson, Sean and Hullman, Jessica and Kay, Matthew},
title = {Uncertainty Displays Using Quantile Dotplots or {CDFs} Improve Transit Decision-Making},
year = {2018},
isbn = {9781450356206},
publisher = {Assoc. for Computing Machinery},
doi = {10.1145/3173574.3173718},
booktitle = {Proc. of the ACM Conference on Human Factors in Computing Systems (CHI)},
pages = {1–12},
numpages = {12},
location = {Montreal QC, Canada},
series = {CHI '18}
}

@article{zacks-tversky-bars-lines,
author = {Zacks, Jeff and Tversky, Barbara},
year = {1999},
month = {12},
pages = {1073-9},
title = {Bars and lines: A study of graphic communication},
volume = {27},
journal = {Memory \& Cognition},
doi = {10.3758/BF03201236}
}

@article{shah-graph-comprehension,
author = {Priti Shah and  Richard E. Mayer and Mary Hegarty},
year = {1999},
month = {12},
pages = {690-702},
title = {Graphs as aids to knowledge construction: Signaling techniques for guiding the process of graph comprehension},
volume = {91},
number = {4},
journal = {Journal of Educational Psychology},
doi = {10.1037/0022-0663.91.4.690}
}

@book{munzner2015visualization,
  author = {Munzner, T.},
  description = {Visualization Analysis and Design - Tamara Munzner - Google Books},
  isbn = {9781498759717},
  publisher = {CRC Press},
  series = {AK Peters Visualization Series},
  title = {Visualization Analysis and Design},
  year = {2015},
doi={10.1201/b17511}
}

@inproceedings{heer-bostock-repr,
author = {Heer, Jeffrey and Bostock, Michael},
title = {Crowdsourcing Graphical Perception: Using Mechanical Turk to Assess Visualization Design},
year = {2010},
isbn = {9781605589299},
publisher = {Assoc. for Computing Machinery},
doi = {10.1145/1753326.1753357},
booktitle = {Proc. of the ACM Conference on Human Factors in Computing Systems (CHI)},
pages = {203–212},
numpages = {10},
location = {Atlanta, Georgia, USA},
series = {CHI '10}
}

@INPROCEEDINGS {bertini-all-chart-not-scatterplot,
author = { Bertini, Enrico and Correll, Michael and Franconeri, Steven },
booktitle = { 2020 IEEE Visualization Conference (VIS) },
title = {{ Why Shouldn’t All Charts Be Scatter Plots? Beyond Precision-Driven Visualizations }},
year = {2020},
volume = {},
ISSN = {},
pages = {206-210},
doi = {10.1109/VIS47514.2020.00048},
publisher = {IEEE Computer Society},
address = {Los Alamitos, CA, USA},
month =Oct}

@book{cleveland_mcgill_2012,
 ISSN = {01621459},
 URL = {http://www.jstor.org/stable/2288400},
 author = {William S. Cleveland and Robert McGill},
 journal = {Journal of the American Statistical Assoc.},
 number = {387},
 pages = {531--554},
 publisher = {[American Statistical Assoc., Taylor \& Francis, Ltd.]},
 title = {Graphical Perception: Theory, Experimentation, and Application to the Development of Graphical Methods},
 urldate = {2023-07-20},
 volume = {79},
 year = {1984}
}

@ARTICLE{borkin-beyond-memorability,
  author={Borkin, Michelle A. and Bylinskii, Zoya and Kim, Nam Wook and Bainbridge, Constance May and Yeh, Chelsea S. and Borkin, Daniel and Pfister, Hanspeter and Oliva, Aude},
  journal={IEEE Transactions on Visualization and Computer Graphics}, 
  title={Beyond Memorability: Visualization Recognition and Recall}, 
  year={2016},
  volume={22},
  number={1},
  pages={519-528},
  doi={10.1109/TVCG.2015.2467732}}

@article{fygenson2023affordances,
  author={Fygenson, Racquel and Franconeri, Steven and Bertini, Enrico},
  journal={IEEE Transactions on Visualization and Computer Graphics}, 
  title={The Arrangement of Marks Impacts Afforded Messages: Ordering, Partitioning, Spacing, and Coloring in Bar Charts}, 
  year={2024},
  volume={30},
  number={1},
  pages={1008-1018},
  doi={10.1109/TVCG.2023.3326590}}

@Manual{r,
     title = {R: A Language and Environment for Statistical Computing},
    author = {{R Core Team}},
    organization = {R Foundation for Statistical Computing},
    address = {Vienna, Austria},
    year = {2021},
    url = {https://www.R-project.org/},
  }

@Misc{qualtrics,
    title = {Qualtrics},
    author = {Qualtrics},
    month = {March},
    year = {2024},
    location = {Provo, Utah, USA},
    url = {https://www.qualtrics.com},
  }

@Misc{prolific,
    title = {Prolific.com},
    author = {Prolific},
    month = {February},
    year = {2025},
    location = {London, UK},
    url = {https://www.prolific.com},
  }

@article{okan-sgl-2019,
author = {Yasmina Okan and Eva Janssen and Mirta Galesic and Erika A. Waters},
title ={Using the Short Graph Literacy Scale to Predict Precursors of Health Behavior Change},
journal = {Medical Decision Making},
volume = {39},
number = {3},
pages = {183-195},
year = {2019},
doi = {10.1177/0272989X19829728}
}

@Manual{r-tidybayes,
    title = {{tidybayes}: Tidy Data and Geoms for {Bayesian} Models},
    author = {Matthew Kay},
    year = {2023},
    note = {R package version 3.0.6},
    url = {http://mjskay.github.io/tidybayes/},
    doi = {10.5281/zenodo.1308151},
}

@Manual{r-tidyr,
  title = {tidyr: Tidy Messy Data},
  author = {Hadley Wickham and Davis Vaughan and Maximilian Girlich},
  year = {2024},
  note = {R package version 1.3.1, https://github.com/tidyverse/tidyr},
  url = {https://tidyr.tidyverse.org},
}

@Article{r-brms,
    title = {{brms}: An {R} Package for {Bayesian} Multilevel Models Using {Stan}},
    author = {Paul-Christian Bürkner},
    journal = {Journal of Statistical Software},
    year = {2017},
    volume = {80},
    number = {1},
    pages = {1--28},
    doi = {10.18637/jss.v080.i01},
    encoding = {UTF-8},
  }

@article{cosmides-tooby-frequency-framing,
	author = {Leda Cosmides and John Tooby},
	doi = {10.1016/0010-0277(95)00664-8},
	issn = {0010-0277},
	journal = {Cognition},
	number = {1},
	pages = {1-73},
	title = {Are humans good intuitive statisticians after all? {R}ethinking some conclusions from the literature on judgment under uncertainty},
	volume = {58},
	year = {1996},
	}

@inproceedings{kay-qdps,
author = {Kay, Matthew and Kola, Tara and Hullman, Jessica R. and Munson, Sean A.},
title = {When (ish) is My Bus? User-centered Visualizations of Uncertainty in Everyday, Mobile Predictive Systems},
year = {2016},
isbn = {9781450333627},
publisher = {Assoc. for Computing Machinery},
doi = {10.1145/2858036.2858558},
booktitle = {Proc. of the ACM Conference on Human Factors in Computing Systems (CHI)},
pages = {5092–5103},
numpages = {12},
location = {San Jose, California, USA},
series = {CHI '16}
}

@article{hullman-hops,
author = {Jessica Hullman and Paul Resnick and Eytan Adar},
title = {Hypothetical Outcome Plots Outperform Error Bars and Violin Plots for Inferences about Reliability of Variable Ordering},
year = {2015},
journal = {PLoS ONE},
volume = {10},
issue = {11},
doi = {10.1371/journal.pone.0142444}
}

@article{cleveland-mcgills-shape-param-graphs,
 ISSN = {01621459},
doi = {10.1080/01621459.1988.10478598},
 author = {William S. Cleveland and Marylyn E. McGill and Robert McGill},
 journal = {Journal of the American Statistical Assoc.},
 number = {402},
 pages = {289--300},
 publisher = {[American Statistical Assoc., Taylor \& Francis, Ltd.]},
 title = {The Shape Parameter of a Two-Variable Graph},
 urldate = {2024-03-23},
 volume = {83},
 year = {1988}
}

@Misc{bayesplot-r,
  title = {Bayesplot: Plotting for Bayesian Models},
  author = {Jonah Gabry and Tristan Mahr},
  year = {2024},
  note = {R package version 1.11.0},
  url = {https://mc-stan.org/bayesplot/},
}

@incollection{Ross-intro-stats,
	address = {Boston},
	author = {Sheldon M. Ross},
	booktitle = {Introductory Statistics (Third Edition)},
	doi = {10.1016/B978-0-12-374388-6.00006-5},
	edition = {Third Edition},
	editor = {Sheldon M. Ross},
	isbn = {978-0-12-374388-6},
	pages = {261-296},
	publisher = {Academic Press},
	title = {CHAPTER 6 - Normal Random Variables},
	year = {2010},
	}

@Manual{ggdist,
    title = {{ggdist}: Visualizations of Distributions and
      Uncertainty},
    author = {Matthew Kay},
    year = {2024},
    note = {R package version 3.3.2},
    url = {https://mjskay.github.io/ggdist/},
    doi = {10.5281/zenodo.3879620},
  }

@ARTICLE{fygenson-padilla-pdf-scaling,
  author={Fygenson, Racquel and Padilla, Lace},
  journal={IEEE Transactions on Visualization and Computer Graphics}, 
  title={Impact of Vertical Scaling on Normal Probability Density Function Plots}, 
  year={2025},
  volume={31},
  number={1},
  pages={984-994},
  doi={10.1109/TVCG.2024.3456396}}

@inproceedings{quadri-doyouseewhatisee,
author = {Quadri, Ghulam Jilani and Wang, Arran Zeyu and Wang, Zhehao and Adorno, Jennifer and Rosen, Paul and Szafir, Danielle Albers},
title = {Do You See What I See? A Qualitative Study Eliciting High-Level Visualization Comprehension},
year = {2024},
isbn = {9798400703300},
publisher = {Assoc. for Computing Machinery},
doi = {10.1145/3613904.3642813},
booktitle = {Proc. of the 2024 CHI Conference on Human Factors in Computing Systems},
articleno = {204},
numpages = {26},
series = {CHI '24}
}

@ARTICLE{xiong-grouping-cues,
author = {Bearfield, Cindy Xiong and Stokes, Chase and Lovett, Andrew and Franconeri, Steven},
title = {What Does the Chart Say? {G}rouping Cues Guide Viewer Comparisons and Conclusions in Bar Charts},
year = {2024},
issue_date = {Aug. 2024},
publisher = {IEEE Educational Activities Department},
address = {USA},
volume = {30},
number = {8},
issn = {1077-2626},
doi = {10.1109/TVCG.2023.3289292},
journal = {IEEE Transactions on Visualization and Computer Graphics},
month = aug,
pages = {5097–5110},
numpages = {14}
}

@ARTICLE {xiong-afford-comparison,
author = {C. Xiong and V. Setlur and B. Bach and E. Koh and K. Lin and S. Franconeri},
journal = {IEEE Transactions on Visualization and Computer Graphics},
title = {Visual Arrangements of Bar Charts Influence Comparisons in Viewer Takeaways},
year = {2022},
volume = {28},
number = {01},
issn = {1941-0506},
pages = {955-965},
doi = {10.1109/TVCG.2021.3114823},
publisher = {IEEE Computer Society},
address = {Los Alamitos, CA, USA},
month = {Jan}
}

@article{carswell-spontaneous-interpretation,
author = {Carswell, C. Melody and Emery, Cathy and Lonon, Andrea M.},
title = {Stimulus complexity and information integration in the spontaneous interpretations of line graphs},
journal = {Applied Cognitive Psychology},
volume = {7},
number = {4},
pages = {341-357},
doi = {10.1002/acp.2350070407},
year = {1993}
}

@article{vessey-cog-fit,
author = {Vessey, Iris},
title = {Cognitive Fit: A Theory-Based Analysis of the Graphs Versus Tables Literature*},
journal = {Decision Sciences},
volume = {22},
number = {2},
pages = {219-240},
doi = {10.1111/j.1540-5915.1991.tb00344.x},
year = {1991}
}

@incollection{gibson-ecological-vis-perception,
  author      = {James J. Gibson},
  title       = {The Theory of Affordances},
  booktitle   = {The Ecol. Approach to Visual Perception},
  publisher   = {Houghton Mifflin},
  year        = {1979},
  pages       = {127-137},
  chapter     = {8},
}

@book{norman-psych-of-everyday-things,
author = {Norman, Donald A.},
title = {The Psychology of Everyday Things},
year = {1988},
isbn = {9780465067091},
publisher = {Basic Books, Inc.},
address = {New York, USA}
}

@article{norman-1999,
	author = {Norman, Donald},
	doi = {10.1145/301153.301168},
	journal = {Interactions},
	month = {05},
	pages = {38-42},
	title = {Affordance, conventions, and design},
	volume = {6},
	year = {1999}}

@book{Kaptelinin_aff_encyclo,
  title        = "Affordances",
  booktitle    = "The Interaction Design Foundation",
  author       = "Kaptelinin, Victor",
  url={https://www.interaction-design.org/literature/book/the-encyclopedia-of-human-computer-interaction-2nd-ed/affordances},
  note         = "Accessed: 2025-1-5",
  language     = "en",
publisher = "Interaction Design Foundation",
year = 2014
}

@inproceedings{hartson-1999-user-actionframework,
	author = {Hartson, H. and Andre, Terence and Williges, Robert and Rens, Linda},
	month = {01},
	pages = {1058-1062},
	title = {The User Action Framework: A Theory-Based Foundation for Inspection and Classification of Usability Problems.},
	year = {1999},
booktitle = {Intl. Journal of Human-Computer Studies},
url={https://dl.acm.org/doi/10.5555/647943.742828}
}

@article{Hartson-2003,
author = {Rex Hartson},
title = {Cognitive, physical, sensory, and functional affordances in interaction design},
journal = {Behaviour \& Information Technology},
volume = {22},
number = {5},
pages = {315--338},
year = {2003},
publisher = {Taylor \& Francis},
doi = {10.1080/01449290310001592587}
}

@ARTICLE{shah-1995-comprehending-line-graphs,
  author={Priti Shah and Patricia A Carpenter},
  journal={Journal of Experimental Psychology: General}, 
  title={Conceptual limitations in comprehending line graphs}, 
  year={1995},
  volume={124},
  number={1},
  pages={43-61},
  doi={10.1037/0096-3445.124.1.43}}

@inbook{ware-info-vis-book,
	author = {Ware, Colin},
	journal = {Information Visualization: Perception for Design: Second Edition},
	month = {04},
	title = {Information Visualization: Perception for Design: Second Edition},
	year = {2004}}

@ARTICLE{isenberg-sys-review-vis-techniques,
  author={Isenberg, Tobias and Isenberg, Petra and Chen, Jian and Sedlmair, Michael and Möller, Torsten},
  journal={IEEE Transactions on Visualization and Computer Graphics}, 
  title={A Systematic Review on the Practice of Evaluating Visualization}, 
  year={2013},
  volume={19},
  number={12},
  pages={2818-2827},
  doi={10.1109/TVCG.2013.126}}

@ARTICLE{livingston-eval-multivar-vis,
  author={Livingston, Mark A. and Decker, Jonathan W. and Ai, Zhuming},
  journal={IEEE Transactions on Visualization and Computer Graphics}, 
  title={Evaluation of Multivariate Visualization on a Multivariate Task}, 
  year={2012},
  volume={18},
  number={12},
  pages={2114-2121},
  doi={10.1109/TVCG.2012.223}}

@article{wagemans-gestalt-perception,
	address = {University of Leuven (KU Leuven), Laboratory of Experimental Psychology, Tiensestraat 102, Box 3711, BE-3000 Leuven, Belgium. johan.wagemans@psy.kuleuven.be},
	author = {Wagemans, Johan and Elder, James H and Kubovy, Michael and Palmer, Stephen E and Peterson, Mary A and Singh, Manish and von der Heydt, R{\"u}diger},
	crdt = {2012/08/01 06:00},
	date = {2012 Nov},
	dcom = {20131104},
	dep = {20120730},
	doi = {10.1037/a0029333},
	edat = {2012/08/01 06:00},
	issn = {1939-1455 (Electronic); 0033-2909 (Print); 0033-2909 (Linking)},
	jid = {0376473},
	journal = {Psychological Bulletin},
	jt = {Psychological bulletin},
	language = {eng},
	lid = {10.1037/a0029333 {$[$}doi{$]$}},
	lr = {20220309},
	mhda = {2013/11/05 06:00},
	mid = {NIHMS390005},
	month = {Nov},
	number = {6},
	own = {NLM},
	pages = {1172--1217},
	phst = {2012/08/01 06:00 {$[$}entrez{$]$}; 2012/08/01 06:00 {$[$}pubmed{$]$}; 2013/11/05 06:00 {$[$}medline{$]$}; 2012/11/01 00:00 {$[$}pmc-release{$]$}},
	pii = {2012-20167-001},
	pl = {United States},
	pmc = {PMC3482144},
	pmcr = {2012/11/01},
	pmid = {22845751},
	pst = {ppublish},
	pt = {Journal Article; Research Support, N.I.H., Extramural; Research Support, Non-U.S. Gov't; Research Support, U.S. Gov't, Non-P.H.S.; Review},
	sb = {IM},
	status = {MEDLINE},
	title = {A century of {Gestalt} psychology in visual perception: {I. P}erceptual grouping and figure-ground organization.},
	volume = {138},
	year = {2012}}

@article{vessey-cog-fit-empirical-study,
 ISSN = {10477047, 15265536},
 URL = {http://www.jstor.org/stable/23010613},
 author = {Iris Vessey and Dennis Galletta},
 journal = {Information Systems Research},
 number = {1},
 pages = {63--84},
 publisher = {INFORMS},
 title = {Cognitive Fit: An Empirical Study of Information Acquisition},
 urldate = {2024-03-18},
 volume = {2},
 year = {1991}
}

@inproceedings{greis-uncertainty,
author = {Greis, Miriam and Joshi, Aditi and Singer, Ken and Schmidt, Albrecht and Machulla, Tonja},
title = {Uncertainty Visualization Influences how Humans Aggregate Discrepant Information},
year = {2018},
isbn = {9781450356206},
publisher = {Assoc. for Computing Machinery},
doi = {10.1145/3173574.3174079},
booktitle = {Proc. of the ACM Conference on Human Factors in Computing Systems (CHI)},
pages = {1–12},
numpages = {12},
location = {Montreal QC, Canada},
series = {CHI '18}
}

@inproceedings{bongshin-lee-task-taxonomy,
author = {Lee, Bongshin and Plaisant, Catherine and Parr, Cynthia Sims and Fekete, Jean-Daniel and Henry, Nathalie},
title = {Task taxonomy for graph visualization},
year = {2006},
isbn = {1595935622},
publisher = {Assoc. for Computing Machinery},
doi = {10.1145/1168149.1168168},
booktitle = {Proc. of the 2006 AVI Workshop on BEyond Time and Errors: Novel Evaluation Methods for Information Visualization},
pages = {1–5},
numpages = {5},
location = {Venice, Italy},
series = {BELIV '06}
}

@article{padilla2018decision,
	author = {Padilla, Lace M. and Creem-Regehr, Sarah H. and Hegarty, Mary and Stefanucci, Jeanine K.},
	date = {2018/07/11},
	doi = {10.1186/s41235-018-0120-9},
	id = {Padilla2018},
	isbn = {2365-7464},
	journal = {Cognitive Research: Principles and Implications},
	number = {1},
	pages = {29},
	title = {Decision making with visualizations: a cognitive framework across disciplines},
	url = {https://doi.org/10.1186/s41235-018-0120-9},
	volume = {3},
	year = {2018}}

@ARTICLE{boy-interactive-vis-aff,
  author={Boy, Jeremy and Eveillard, Louis and Detienne, Françoise and Fekete, Jean-Daniel},
  journal={IEEE Transactions on Visualization and Computer Graphics}, 
  title={Suggested Interactivity: Seeking Perceived Affordances for Information Visualization}, 
  year={2016},
  volume={22},
  number={1},
  pages={639-648},
  doi={10.1109/TVCG.2015.2467201}}

@inproceedings{taher2015physicalaffbarchart,
author = {Taher, Faisal and Hardy, John and Karnik, Abhijit and Weichel, Christian and Jansen, Yvonne and Hornb\ae{}k, Kasper and Alexander, Jason},
title = {Exploring Interactions with Physically Dynamic Bar Charts},
year = {2015},
isbn = {9781450331456},
publisher = {Assoc. for Computing Machinery},
doi = {10.1145/2702123.2702604},
booktitle = {Proc. of the ACM Conference on Human Factors in Computing Systems (CHI)},
pages = {3237–3246},
numpages = {10},
series = {CHI '15}
}

@ARTICLE{correll-prob-density2019,
  author={Correll, Michael and Li, Mingwei and Kindlmann, Gordon and Scheidegger, Carlos},
  journal={IEEE Transactions on Visualization and Computer Graphics}, 
  title={Looks Good To Me: Visualizations As Sanity Checks}, 
  year={2019},
  volume={25},
  number={1},
  pages={830-839},
  doi={10.1109/TVCG.2018.2864907}}

@article{Nisbett1977-NISTMT,
	author = {Richard E. Nisbett and Timothy D. Wilson},
	doi = {10.1037/0033-295x.84.3.231},
	journal = {Psychological Review},
	number = {3},
	pages = {231--59},
	title = {Telling More Than We Can Know: Verbal Reports on Mental Processes},
	volume = {84},
	year = {1977}
}

@article{gauvrit-eqprob,
	author = {Gauvrit, Nicolas and Morsanyi, Kinga},
	journal = {Adv Cogn Psychol},
	number = {4},
	pages = {119--130},
	title = {The equiprobability bias from a mathematical and psychological perspective.},
	volume = {10},
	year = {2014},
doi={10.5709/acp-0163-9}}

@article{fygenson-cog-affordance-framework,
  author={Fygenson, Racquel and Padilla, Lace and Bertini, Enrico},
  journal={IEEE Transactions on Visualization and Computer Graphics}, 
  title={Cognitive Affordances in Visualization: Related Constructs, Design Factors, and Framework}, 
  year={2025},
  volume={31},
  number={12},
  pages={10624-10639},
  doi={10.1109/TVCG.2025.3610803}}

@article{kruschke2021bayesianreporting,
	author = {Kruschke, John K. and Liddell, Torrin M.},
	date = {2018/02/01},
	doi = {10.3758/s13423-017-1272-1},
	id = {Kruschke2018},
	isbn = {1531-5320},
	journal = {Psychonomic Bulletin \& Review},
	number = {1},
	pages = {155--177},
	title = {Bayesian data analysis for newcomers},
	url = {https://doi.org/10.3758/s13423-017-1272-1},
	volume = {25},
	year = {2018}}

\end{document}